\documentclass{aa}

\usepackage{rotate,psfig}

\begin{document}

\thesaurus{03       
              (11.17.3;                 
               11.01.2;                 
               13.18.1;                 
               12.03.3;                 
                               )} 

\title{ The ``angular size - redshift'' relation for compact radio
structures in quasars and radio galaxies \thanks{Table 1 is available
in electronic form at the CDS via anonymous ftp to {\tt
cdsarc.u-strasbg.fr (130.79.128.5)} or via {\tt
http://cdsweb.u-strasbg.fr/Abstract.html}}  }

\author{L.~I.~Gurvits\inst{1,2}
\and
K.~I.~Kellermann\inst{3}
\and
S.~Frey\inst{4,1}
}

\offprints{L.~I.~Gurvits, 1st address, {\tt lgurvits@jive.nfra.nl}}

\institute{Joint Institute for VLBI in Europe, P.O. Box 2, 7991 PD
Dwingeloo, The Netherlands
\and
Astro Space Center, P.N.Lebedev Physical Institute, Moscow 117924,
Russia 
\and
National Radio Astronomy Observatory, 520 Edgemont Road,
Charlottesville, 
Virginia 22903-2475, USA 
\and
F\"OMI Satellite Geodetic Observatory, P.O. Box 546, H--1373 Budapest,
Hungary
}

\date{Received 24 Jun 1998 /  
      Accepted 05 Oct 1998}

\titlerunning{The ``angular size -- redshift'' relation for compact
radio structures ...}
\authorrunning{L.I.Gurvits et al.}
\maketitle

\vspace{0.5cm}

\begin{abstract}

We discuss the ``angular size -- redshift'' relation for compact radio
sources distributed over a wide range of redshifts $0.011 \le z \le
4.72$.  Our study is based on a sample of 330 5 GHz VLBI contour maps
taken from the literature. Unlike extended source samples, the
``angular size -- redshift'' relation for compact radio sources appears
consistent with the predictions of standard Friedmann world models with
$q_{\circ} \simeq 0.5$ without the need to consider evolutionary or
selection effects due to a ``linear size -- luminosity'' dependence.
By confining our analysis to sources having a spectral index, $-0.38
\le \alpha \le 0.18$, and a total radio luminosity, $Lh^{2} \ge
10^{26}$ W/Hz ($H_{\circ} = 100\,h$ km\,s$^{-1}$\,Mpc$^{-1}$,
$q_{\circ} = 0.5$ used as a numerical example), we are able to restrict
the dispersion in the ``angular size -- redshift'' relation. The best
fitting regression analysis in the framework of the
Friedmann-Robertson-Walker model gives the value of the deceleration
parameter $q_{\circ} = 0.21 \pm 0.30$ if there are no evolutionary or
selection effects due to a ``linear size -- luminosity'', ``linear size
-- redshift'' or ``linear size -- spectral index'' dependence.


\end{abstract}

\section{Introduction}

Classical tests of cosmological world models using the observed
dependence of the angular size of galaxies or kiloparsec-scale radio
sources have been inconclusive.  At optical wavelengths, observational
uncertainties at large redshift are large due to the small size of a
galactic disk, seeing, the difficulty in defining a true metric rod,
and possible evolutionary effects (e.g.  Sandage 1988).  At radio
wavelengths, the separation of the lobes of extended double radio
sources may be determined with great accuracy even at large redshift,
but the interpretation of the ``angular size -- redshift'' ($\theta -
z$) relation for double radio sources has been obscured by possible
selection and evolutionary effects.  The observed $\theta - z$ relation
for double radio sources appears to follow a simple $1/z$ law even at
high redshift, in apparent contradiction to any simple
Friedmann-Robertson-Walker (FRW) model without evolution (e.g. Kapahi
1989).  Most researchers interpret the observed $\theta - z$ diagram
for double radio sources as evidence for a decrease in linear size with
redshift (Kapahi 1987, Barthel and Miley 1988, Neeser et al. 1995).
However, Singal (1993) and Nilsson et al.  (1993) consider that the
observed departure from the FRW curves is due to an inverse ``linear
size -- luminosity'' correlation which preferentially selects the
smaller (high luminosity) sources at high redshifts. It is curious,
however, that these selection or evolutionary effects apparently
combine with cosmological effects to give the simple observed $1/z$
relation.

More recently, Buchalter et al. (1998) have studied a sample of 103
double lobed quasars with $z > 0.3$ using the VLA at 20 cm in its
B-configuration. In contrast to the $1/z$ ``angular size -- redshift''
relation found for double lobed radio sources by other workers,
Buchalter et al. find no change in apparent angular size in the range
of $1.0 \la z \la 2.7$,  consistent with FRW models without significant
evolution.  But, it is not clear to what extent their results are
affected by the limited range of angular size, between 12 and 120
arcseconds, which can be observed with the VLA in the B-configuration
at 20 cm.

The size of extended double lobe source whose linear extent is
typically hundreds of kiloparsecs, may depend on the systematic changes
in the properties of the intergalactic medium with $z$.  Moreover, high
redshift extended sources have ages which are comparable to the age of
the Universe, and so evolutionary effects are not unexpected.  Compact
radio jets associated with quasars and AGN, by contrast, are typically
less than a hundred parsec in extent. Their morphology and kinematics
probably depends more on the nature of the ``central engine'' than on
the surrounding intergalactic medium. The ``central engine'' itself is
thought to be controlled by a limited number of physical parameters,
such as the mass of central black hole, the strength of magnetic field,
the accretion rate, and, possibly, the angular momentum. This central
region may be ``standard'' for sources in which these parameters are
confined within restricted ranges.  Also, because the compact radio
jets have typical ages of only some tens of years, they are young
compared to the age of the Universe, at any reasonable redshift.
Therefore, compact radio sources may offer an evolution free sample to
test world models over a wide range of redshift.

However, the size of compact radio jets is not unambiguously defined
and depends on the frequency of observation as well as on resolution.
Moreover, differences in the spectral index between the core and jet
components may introduce a K-like correction which can be important for
high redshift sources, as this may introduce an apparent ``linear size
-- redshift'' dependence even in the absence of evolution (Kellermann
1993). Frey et al. (1997) have shown that this is likely to be a weak
dependence, however more detailed images at various frequencies with 
matched resolution be needed to verify the importance of any K-like
correction.

In several previous studies we have reported on the observed $\theta -
z$ relation for compact radio sources. Kellermann (1993) studied a
sample of 79 quasars and AGN's which had been observed with VLBI at 5
GHz and which have a 5 GHz luminosity greater than $10^{24}$ W/Hz.
There are only a few sources at low redshift which meet the luminosity
restriction, but these are consistent with a $1/z$ relation,
characteristic of the Euclidean geometry which describes the local
Universe. The Kellermann (1993) sample already includes all sources
with luminosity  greater than $10^{24}$~W/Hz at redshifts less than a
few tenths, and further surveys down to fainter flux density limits
will not find any additional sources which satisfy the luminosity
criteria. The important point of the Kellermann (1993) paper was that
for redshifts in the range $0.5 < z < 3$, the angular size appears to
be essentially independent of redshift, in contrast to the $\theta - z$
relation for powerful extended sources which continues its apparent
Euclidean form out to large redshifts. Kellermann noted that the
observed form of the $\theta - z$ relation for the compact source
sample was qualitatively consistent with a standard FRW cosmology with
$\Omega = 1$ without the need to appeal to arguments based on size
evolution or a ``linear size -- luminosity'' dependence. A more
rigorous quantitative analysis of this data by Stepanas and Saha (1995)
find a best fit of $q_{\circ} = 2.6^{+2.1}_{-2.2}$ with the 90\%
confidence, and they exclude the simple $\theta \propto 1/z$ relation
at the 99\% confidence level. Using the same data, Kayser (1995) has
applied a Kolmogorov-Smirnov test to compare the linear sizes of high
($z > 0.75$) and low ($z < 0.75$) redshift compact radio sources for
different cosmological models and also concludes that the available
data allow models with a wide range of $\Omega$ and the cosmological
constant, $\Lambda$.

In a separate investigation, Gurvits (1993) used two point VLBI
visibility data obtained at 13 cm (Preston et al.  1985) for 337
sources in order to show qualitatively that the observed data suggests
$q_{\circ} \le 0.5$. A four-parameter regression analysis of the same
sample gave a value of $q_{\circ} = 0.16 \pm 0.71$ (Gurvits 1994).  The
same analysis also gave estimates of the dependence of the apparent
angular sizes of compact sources on their luminosity and emitting
frequency.

More recently, Wilkinson et al. (1998) have reported on the $\theta -
z$ relation for sources taken from the Caltech-Jodrell Bank VLBI CJF
sample of 160 flat spectrum radio sources (Taylor et al. 1996 and
references therein). As in the studies of Kellermann (1993) and Gurvits
(1994), Wilkinson et al. find no dependence of angular size on redshift
for sources with $0.5 < z < 3$, but conclude that uncertainties in
defining the angular size of complex jets, in the K-like correction, in
a possible ``size -- luminosity'' dependence, in the effects of
orientation, as well as in possible size or luminosity evolution
restrict the usefulness of compact sources to accurately constrain the
value of $q_{\circ}$.  Dabrowski, Lasenby, and Saunders (1995), as
well, have pointed out the difficulty in obtaining a meaningful
constraint on $\Omega$ due to the effects of relativistic beaming in
limited source samples.

Krauss and Schramm (1993) and Stelmach (1994) have pointed out that if
evolutionary effects can be ruled out, then the form of the ``angular
size -- redshift'' relation can put significant limits on the value of
the cosmological constant, $\Lambda$, as well as on $\Omega$.  Jackson
and Dodgson (1996) have shown that while the data presented by
Kellermann (1993) are consistent with $\Omega_{\circ} \sim 1$ and
$\Lambda_{\circ} = 0$, since there is not a well defined minimum in the
$\theta -z$ dependence, equally good fits to the data are obtained with
smaller values of matter density described by $\Omega_{\circ} < 1$, and
negative values of $-6 \la \Lambda_{\circ} \la -2 $.

With the aim of better restricting the allowable range of cosmological
parameters, we have compiled a new larger sample of sources than used
by Kellermann (1993) or by Wilkinson et al. (1998) but with more
complete structural data than used by Gurvits (1993, 1994). We note,
that the sample discussed here is inhomogeneous as it is based on VLBI
images published by various authors using a variety of antenna
configurations and different techniques for image reconstruction.  In
Section 2 we discuss the definition of our sample, and in the following
sections we discuss the apparent ``angular size -- redshift''
dependence.

\section{ The source sample }

The new list contains all sources found in the literature which were
imaged with VLBI at 5 GHz with a nominal resolution of about 1.5 mas in
the east-west direction and with a dynamic range of at least 100.  The
list includes the all-sky set of 79 sources discussed by Kellermann
(1993), but enhanced by more recently published work, mostly by the
Caltech--Jodrell Bank group (Xu et al. 1995, Henstock et al. 1995,
Taylor et al. 1994, 1996) of sources at declination greater than 35
degrees.  We also included in the sample a number of sources published
by other authors and our own recent observations of quasars with
measured redshifts greater than 3 (Frey et al. 1997, Paragi et al.
1998).

Our sample differs from the compilation of Wilkinson et al. (1998)
primarily in that it includes a number of relatively strong sources at
declinations south of +35 degrees, and other sources not presented in
the CJF sample. By including sources outside the range of CJF
declinations we are able to better sample the sparsely populated low
redshift (Euclidean) part of the $\theta - z$ diagram not included in
the CJF sample. The observations of Gurvits et al. (1992, 1994), Frey
et al. (1997) and Paragi et al. (1998) were made in an attempt to
better sample the high redshift part of the $\theta - z$ diagram which
is particularly sensitive to the value of $q_{\circ}$.  In particular,
the unambiguous detection of an increase in angular size at the highest
redshifts would indicate a value of $q_{\circ} > 0$.  The increase in
the size of our present sample comes at the expense of homogeneity and
the need to use published VLBI contour maps instead of the primary
data.  We have attempted to minimize the effect of these
inhomogeneities by using the following criteria.

As in Kellermann (1993), we define the characteristic angular size of
each source as the distance between the strongest component, which we
refer to as the core, and the most distant component which has a peak
brightness greater than or equal to 2\% of the peak brightness of the
core. For sources which are slightly resolved or unresolved, we adopted
the following procedure. We assume that sources are one dimensional.
For sources which appear resolved in at least one direction, we
estimated the distance of a secondary component from the core, or its
upper limit, from the published contours. If the source was not
resolved, we took the size of the major axis of the synthesized beam
(FWHM) as an upper limit to the size, unless there was additional
information which indicated that the source structure axis lies along a
specific direction different from the direction of major axis. The
latter applies to those sources which show extensions in a particular
direction, including an extension along the minor axis of the
synthesized beam. In this case we estimated the upper limit of the size
as the size of the beam along the direction of extension. Thus, our
approach to measuring source size allows for four different cases:

{\bf Case C:} the distance between the core and a 2\%-component;

{\bf Case J:} an upper limit of the size measured as the size of the
synthesized beam along the direction of apparent extension, most likely
-- a jet;

{\bf Case L:} an upper limit of the size measured along the major axis
of the synthesized beam;

{\bf Case S:} an upper limit of the size measured along the minor axis
of the synthesized beam.

For those sources where multi epoch VLBI images are available, we have
used the most recent epoch that meets our criteria of sensitivity and
dynamic range. Finally, we excluded from our analysis all unresolved
sources if the major axis of the primary beam exceeded 2.2 mas (i.e.
all cases L with too large a synthesized beam) and sources which
are known to be gravitationally lensed.

The resulting sample of 330 sources is presented in
Table~\ref{TSou1}\footnote{Table 1 is available in electronic form at
the CDS via anonymous ftp to {\tt cdsarc.u-strasbg.fr (130.79.128.5)}
or via {\tt http://cdsweb.u-strasbg.fr/Abstract.html}}, where we show
the IAU source designation and alternative name in columns 1 and 2. The
redshift and optical counterpart are given in columns 3 and 4.  Columns
5, 6 and 7 give the flux density at 6 and 20 cm (or a footnote for
alternative wavelength) and the two-point spectral index, $\alpha$ ($S
\propto \nu ^{\alpha}$), respectively.  Columns 8 and 9 give the
angular size (or its upper limit) and the one-letter structure code as
explained above.  In columns 10--13, we list references for redshift,
flux densities at 6 and 20 cm, and for the VLBI image respectively.

\section{ Properties of the sample }

\begin{figure*}[h]
\centerline{
\rotate[r]
{
\psfig{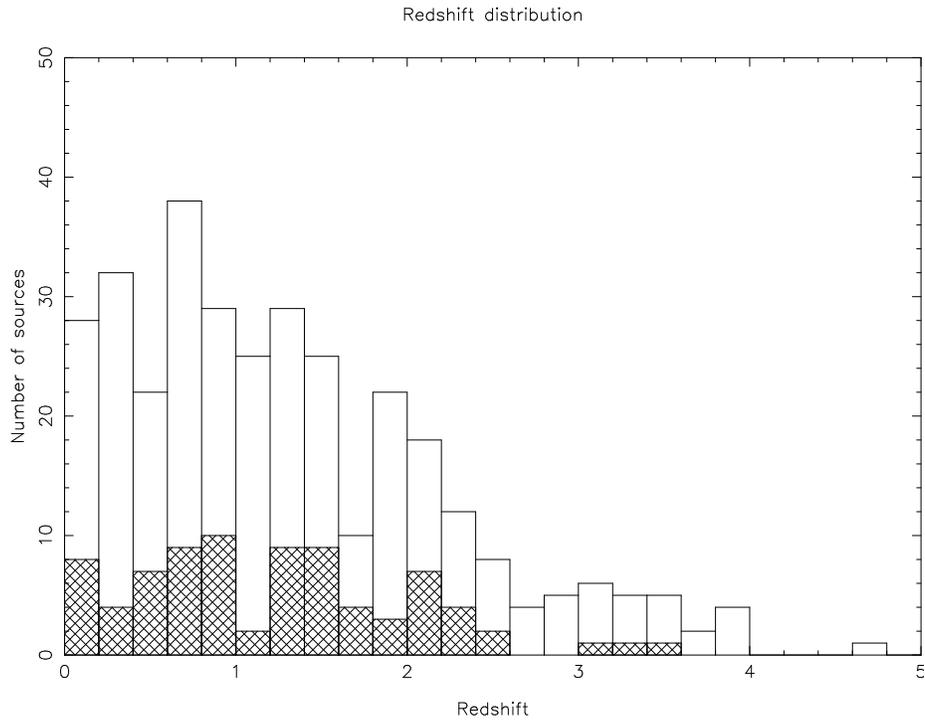}}
}
\caption{ Redshift distribution for the full sample of 330 sources.
Shaded part of the histogram represents the 79 sources from Kellermann
(1993).
        }
\label{zhis}
\end{figure*}

The distribution of redshifts for the sources in our sample is shown in
Fig.~\ref{zhis}. The 79 sources used in the analysis of Kellermann (1993) are
shown shaded.

\begin{figure*}[h]
\centerline{
\rotate[r]
{
\psfig{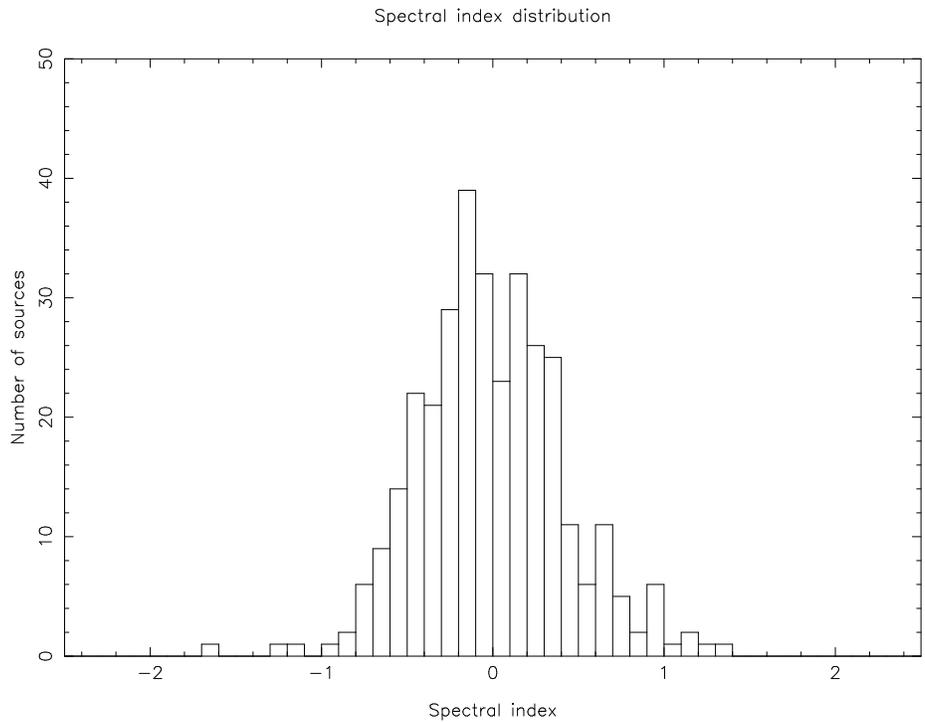}}
}
\caption{ Spectral index distribution for the sample of 330 sources ($S
\propto \nu ^{\alpha}$).
     }
\label{spihis}
\end{figure*}

The histogram of the spectral index distribution is shown in
Fig.~\ref{spihis}.  We note, that we have used the value of spectral
index as calculated, in most cases, from measurements of total flux
density on arcsecond or larger angular scales although our discussion
of angular dimensions is based on milliarcsecond-scale structures,
which account for only part of the total flux. For most of the sources,
this distinction is not important as nearly all of the flux density in
sources of interest is contained in the compact component.  Where
relevant, such as for Cyg A, we specifically used the flux density of
the core component. In a few special cases marked in column 6, when
flux density at 20 cm was not available, the spectral index was
calculated between 6 cm and another longer wavelength as explained in
the footnotes.

Fig.~\ref{lumi} shows the luminosity of all the sources in our sample as a
function of their redshift.  (Throughout this paper we use $H_{\circ}
= 100\,h {\rm \,\, km \, s^{-1} \, Mpc^{-1}}$ and a deceleration
parameter  $q_{\circ} = 0.5$ to calculate the luminosity).  The shape of
the luminosity -- redshift diagram and the narrow dispersion simply
reflects the fact that our sample, although compiled on an ad-hoc basis
from the literature and based upon various selection criteria, is
basically a flux-limited sample.

\section{ Properties of angular size }

In Fig.~\ref{thez330}, we plot the measured angular size against
redshift for all 330 sources in our sample. For the well resolved
sources, the procedure of measuring $\theta$ gives an unambiguous
estimate of a metric size.  But, for sources with maximum dimensions
comparable to or smaller than the beam size, there are large
uncertainties. For this reason, and to minimize the influence on our
analysis of the few sources with extremely large dimensions, we have
chosen to bin the data and to examine the change in median angular size
with redshift. This allows us to treat equally true metric sizes of
resolved sources and upper limits of sizes for slightly resolved or
unresolved ones. Fig.~\ref{mtz18} shows the binned data of median
angular size plotted against redshift for the same data.  (Here and
throughout this paper we use nearly equally populated bins, which
number is close to $\sqrt N$, where $N$ is the size of the sample.) As
found in previous studies, for $z \ge 0.5$, the median angular size is
nearly independent of redshift. In this figure, as an example we show a
family of curves for a standard rod in various world models. We note,
that none of these curves represent the best fit discussed below.

\subsection { ``Angular size -- luminosity'' and \protect\\ ``angular
size -- spectral index'' relations }

\begin{figure*}[h]
\centerline{
\rotate[r]
{
\psfig{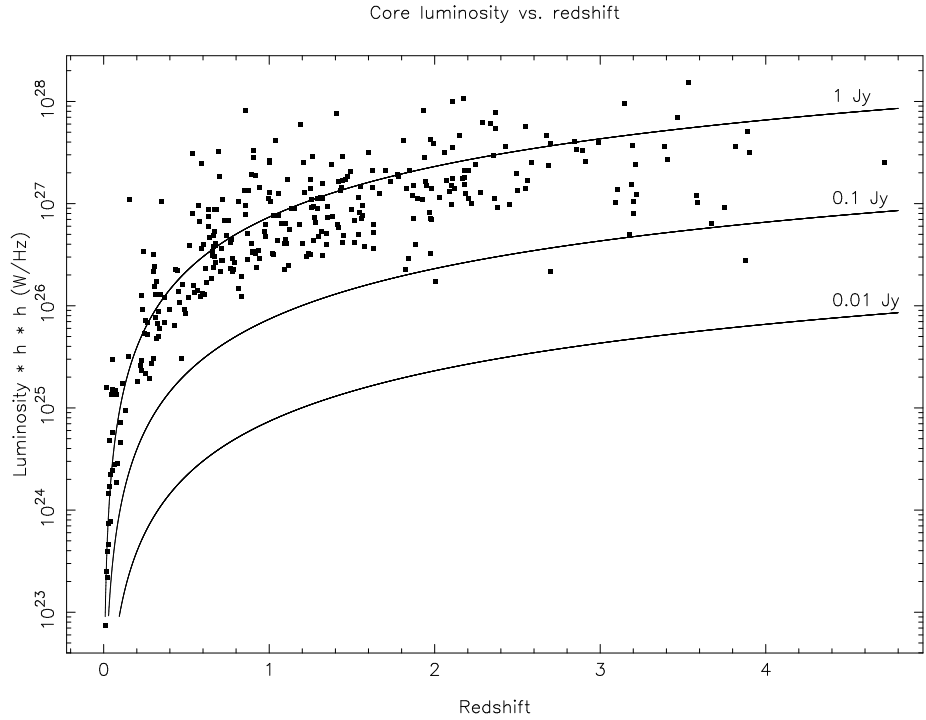}}
}
\caption{ Luminosity as a function of redshift for the full sample of
330 sources (shown by filled squares) calculated with $H_{\circ} =
100\,h$ km\,s$^{-1}$\,Mpc$^{-1}$ and $q_{\circ} = 0.5$ as a numerical
example.  The latter values have not been used in the regression
analysis described in subsection 4.2. The solid lines show luminosities
of sources with  flux density of 1, 0.1 and 0.01 Jy, calculated under
assumption that the spectral index $\alpha = 0$.
        }
\label{lumi}
\end{figure*}

\begin{figure*}[h]
\centerline{
{
\psfig{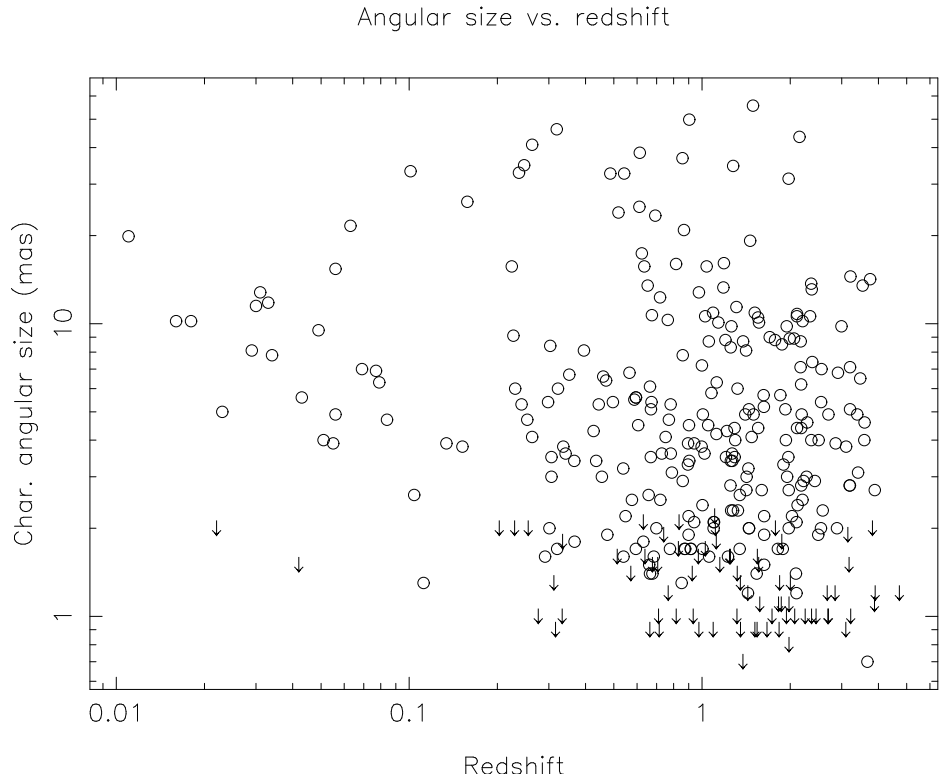}}
}
\caption{ ``Angular size -- redshift'' diagram for the sample of 330
sources. Measured sizes are shown with empty circles, upper limits --
with arrows.
        }
\label{thez330}
\end{figure*}

\begin{figure*}[h]
\centerline{
{
\psfig{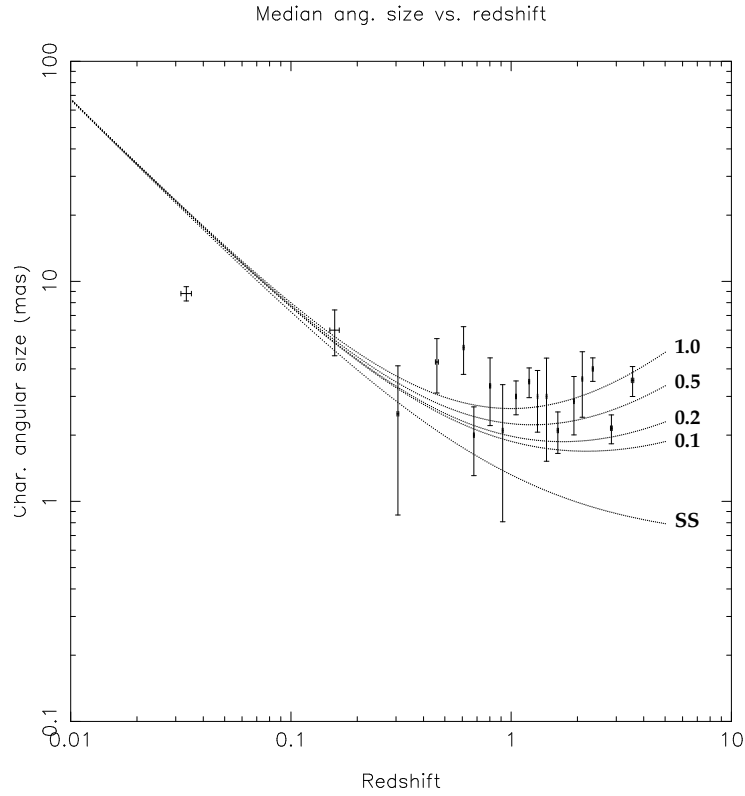}}
}
\caption{ Median angular size versus redshift. The full length of error
bars here and in other figures corresponds to 1$\sigma$. The solid
lines correspond to the linear size parameter $lh = 9.6$ pc, the
Steady-state model (SS) and models of a homogeneous, isotropic Universe
with $\Lambda = 0$ and values of $q_{\circ} = 1.0, \, 0.5, \,0.2, \,
0.1$ (as marked on the plot). Data are binned into 18 bins nearly
equally populated (18--19 sources per bin).
     }
\label{mtz18}
\end{figure*}

\begin{figure*}[h]
\centerline{
\rotate[r]
{
\psfig{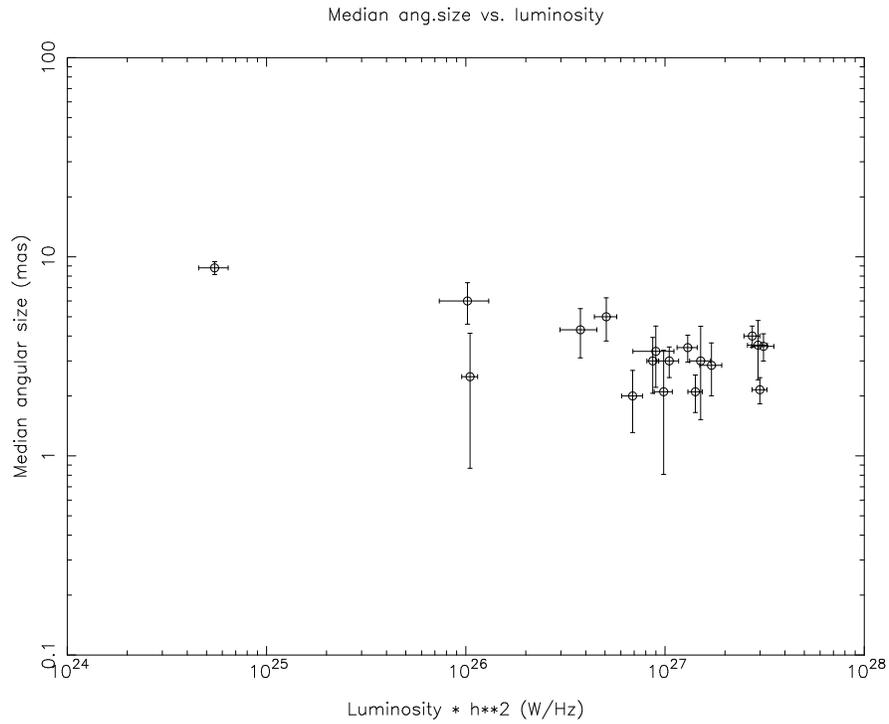}}
}
\caption{ Median angular size versus luminosity for 18 bins in redshift
space (the same as in Fig.~\ref{mtz18}).
     }
\label{ml18}
\end{figure*}

As it is clear from Fig.~\ref{lumi}, our sample contains sources with
luminosity ranging over more than 4 orders of magnitude.
Fig.~\ref{ml18} shows the relation between median angular size  and
luminosity (the same binning in redshift space as in
Fig.~\ref{mtz18}).

\begin{figure*}[h]
\centerline{
\rotate[r]
{
\psfig{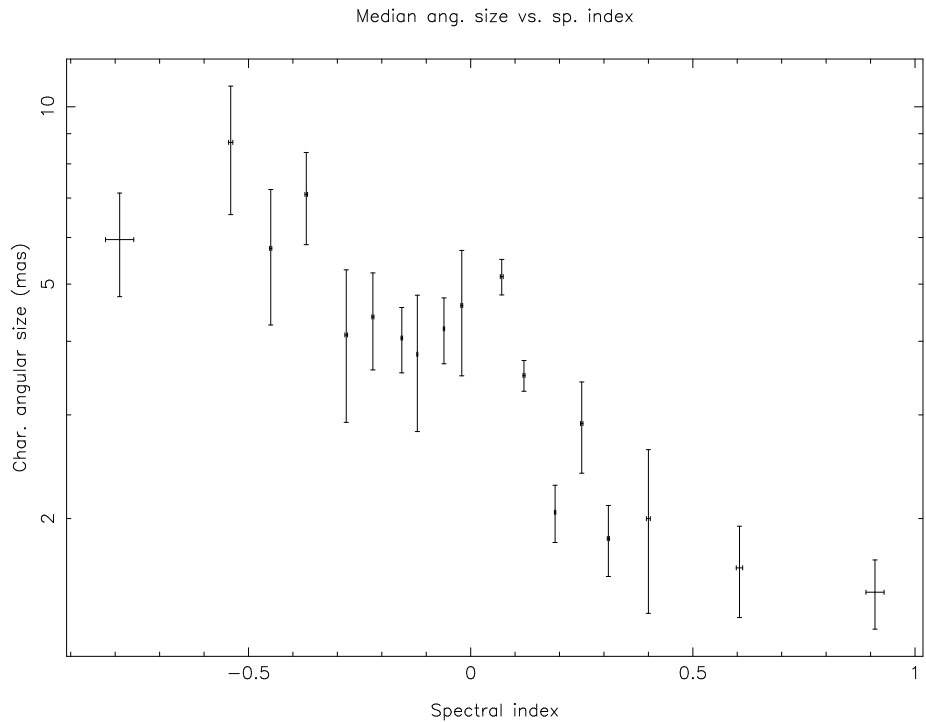}}
}
\caption{ Median angular size as a function of spectral index (the same
binning in redshift space as in Fig.~\ref{mtz18}). 
        }
\label{mspi18}
\end{figure*}

\begin{figure*}[h]
\centerline{
\rotate[r]
{
\psfig{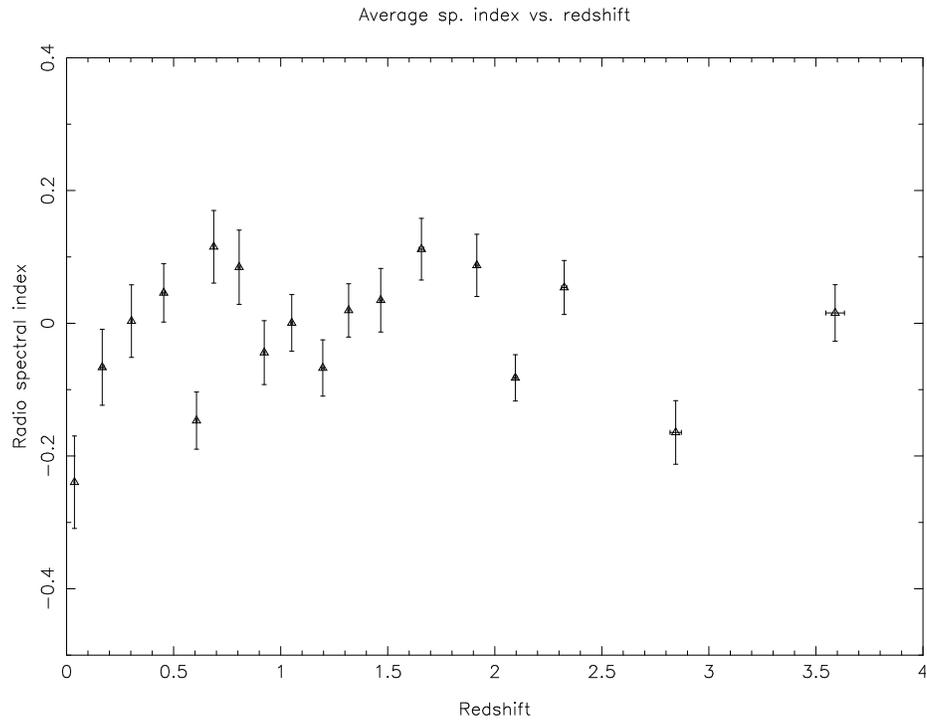}}
}
\caption{ Spectral index versus redshift (the same binning in redshift
space as in Fig.~\ref{mtz18}).
        }
\label{spiz18}
\end{figure*}

Fig.~\ref{mspi18} shows the dependence of the angular size on spectral
index. As expected from simple consideration of self absorption
arguments, sources with flat and inverted spectra should, on average,
have smaller sizes. Fig.~\ref{mspi18} qualitatively confirms this
expectation. It may also indicate a presence of one or several
selection effects. However, as illustrated by Fig.~\ref{spiz18}, we do
not find an evidence on systematic correlation between $\alpha$ and
$z$, which might be responsible for the appearance of the ``$\theta -
\alpha$'' dependence, shown in Fig.  7. The only possible exception
could correspond to the lowest redshift bin. However, this bin
represents sources of considerably lower luminosity (cf.
Fig.~\ref{lumi}), which could differ intrinsically from their higher
redshift counterparts.

\subsection{ Toward estimating cosmological parameters from the
\protect \\ $\theta - z$ relation }

The $\theta-z$ relation based on the data described here is
qualitatively consistent with $0 \le q_{\circ} \le 1$ and $\Lambda = 0$
without the need to introduce evolutionary effects.  The new data, in
agreement with presented earlier by Kellermann (1993), Gurvits (1994)
and Wilkinson et al. (1998), do not show clear evidence for an angular
size minimum near $z=1.25$ as expected for models with $\Omega = 1$ and
$\Lambda = 0$.  The near asymptotic slope of the $\theta - z$ relation
is more characteristic of models with $\Omega < 1$ and allows values of
$\Lambda \ne 0$.

These results are, however, based on very inhomogeneous data obtained
by many different observers using different instruments and imaging
techniques.  New VLBI observations now in progress will improve the
accuracy of the observed $\theta - z$ relation as it will provide a
uniform data set for analysis using the $uv$-data and images rather
than published contour maps.

With all the reservations discussed above, as an example of a
cosmological test with the $\theta - z$ relation on milliarcsecond
scale, we consider a multi-parameter regression analysis as described
by Gurvits (1994) with modifications made by Frey (1998). It is based
on the following phenomenological expression

\begin{equation}
\theta \equiv l_{m}\,D^{-1}(z) \propto lh\,L^{\beta}\,(1+z)^{n}\,D^{-1}(z)
\,\,\,\,  ,
\end{equation}

\noindent where $l_{m}$ is the metric linear size, $D$ is the angular
size distance, $lh$ is the linear size scaling factor, $L$ is the
source luminosity. Parameters $\beta$ and $n$ represent the dependence
of the linear size on the source luminosity and redshift,
respectively.  For a homogeneous, isotropic Universe ($q_{\circ} > 0$)
with the cosmological constant, $\Lambda = 0$, $D(z)$ is given by the
usual expression

\begin{equation}
D(z) =
\frac{q_o z + (q_o-1)\left(\sqrt{1+2q_{o}z} - 1 \right)}
{q_{o}^{2}(1+z)^2} \,\,\,\,  .
\end{equation}

The regression model (Gurvits 1994, Frey 1998) allows us to fit the
$\theta - z$ relation with four free parameters, the linear size
scaling factor $lh$, the deceleration parameter $q_{0}$ and two
parameters related to the physics of compact radio emitting regions,
$\beta$ and $n$. The value of $n$, in turn, could in principle
represent three different physical dependences: {\it (i)} a
cosmological evolution of the linear size; {\it (ii)} a dependence of
the linear size on the emitted frequency; and {\it (iii)} an impact of
sources broadening due to scattering in the propagation medium. The
latter effect is not important for our sample with the lowest emitted
frequency of 5 GHz (which corresponds to $z=0$). The distinction
between the former two effects is beyond the scope of this paper and
will require multifrequency $\theta - z$ tests.

\begin{figure*}[h]
\centerline{
\rotate[r]
{
\psfig{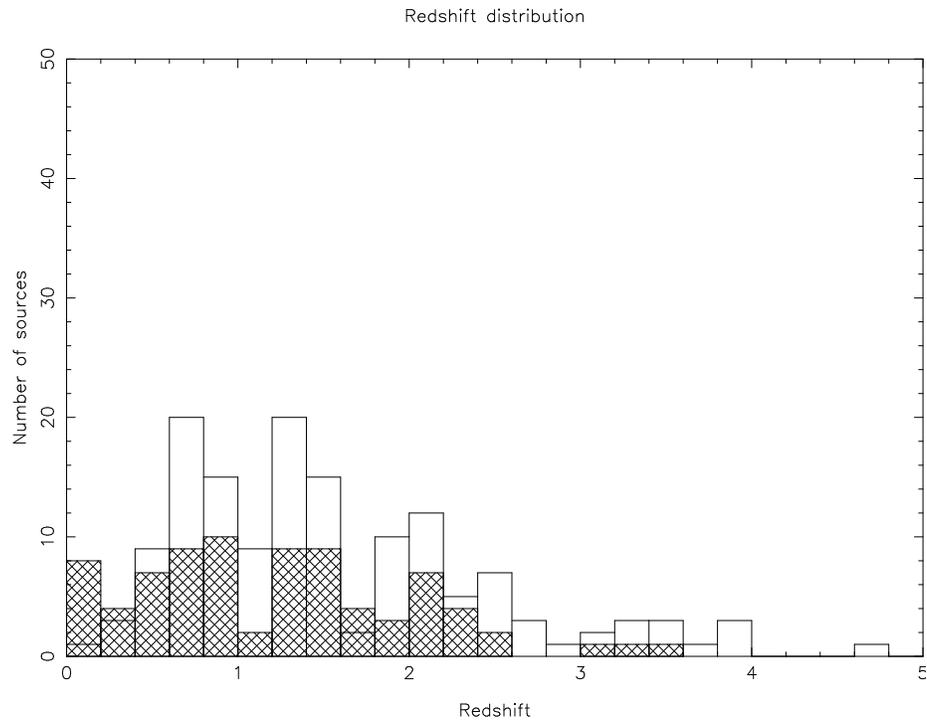}}
}
\caption{ Redshift distribution for 145 sources with $L \ge 10^{26}$
W/Hz and $-0.38 \le \alpha \le 0.18$. Shaded part of the histogram
represents the 79 sources from Kellermann (1993).
        }
\label{zhi145}
\end{figure*}

\begin{figure*}[h]
\centerline{
{
\psfig{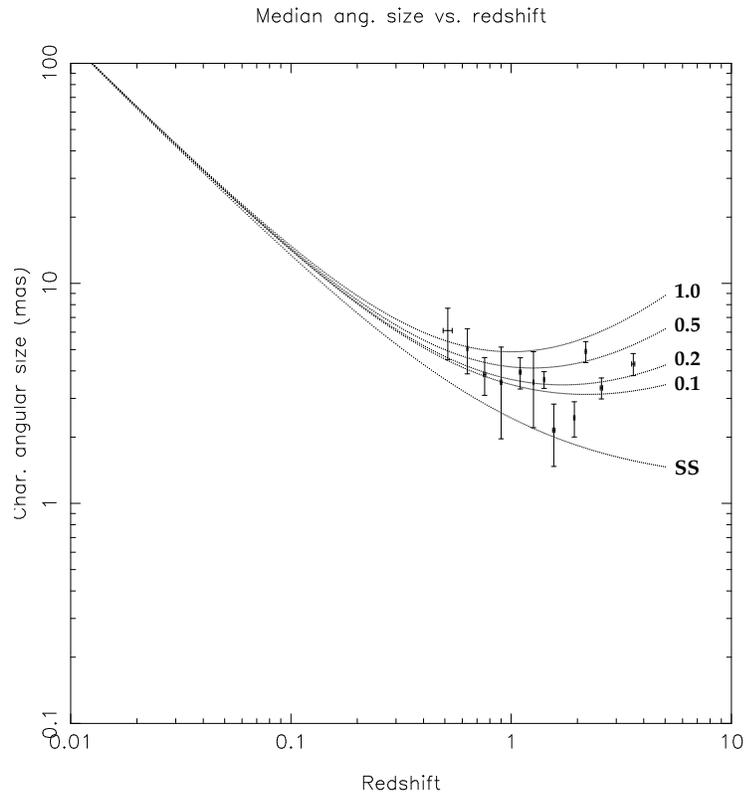}}
}
\caption{ Median angular size versus redshift for 145 sources  (binned
into 12 bins, 12--13 sources per bin) with $-0.38 \le \alpha  \le 0.18$
and $L \ge 10^{26}$ W/Hz. The solid lines correspond to the linear size
parameter $lh = 22.7$ pc, the Steady-state model (SS) and models of a
homogeneous, isotropic Universe with $\Lambda = 0$ and various shown
values of $q_{\circ}$.  None of the solid lines represents the best
fit.
   }
\label{mtz12}
\end{figure*}

To minimize any possible dependence of linear size on luminosity, we
restrict the regression analysis to sources with $Lh^{2} \ge 10^{26}$
W/Hz. Furthermore, as is evident from Fig.~\ref{mspi18}, there is an
obvious dependence of angular size on spectral index. In order to
minimize this effect on the regression, we choose only those sources
which form a flat segment of the $\theta - \alpha$ diagram $-0.38 \le
\alpha \le 0.18$ (Fig.~\ref{mspi18}). This selection criterion also
partially excludes from the analysis the lowest redshift bin which
represents the highest deviation on the $\alpha - z$ diagram
(Fig.~\ref{spiz18}).  By restricting the range of spectral indices, we
are able to further restrict the dispersion in intrinsic size in our
analysis.  Specifically, we exclude many of the relatively large
compact steep spectrum sources and most compact inverted spectrum
sources.  There are 145 sources which meet these criteria, their
distribution in redshift space is shown in Fig.~\ref{zhi145}, and the
median angular sizes versus redshift is shown in Fig.~\ref{mtz12}.

As an example, we apply the four parameter regression model for median
values of this sub-sample grouped into 12 redshift bins. The best fit
values and corresponding 1$\sigma$ errors are: $lh = 23.8 \pm 17.0$ pc,
$\beta = 0.37 \pm 0.27$, $n = -0.58 \pm 1.0$, and $q_{0} = 0.33 \pm
0.11$. This result is in qualitative agreement with similar estimates
obtained for an independent sample of sources and different technique
of measuring their angular sizes (Gurvits 1994).

In Table~\ref{Tregr}, we show the results of regression modeling of the
same binned sub-sample for the two parameters, $lh$ and $q_{0}$, for
different fixed values of $\beta$ and $n$. The ranges for $\beta$ and
$n$ shown do not require a substantial evolution of linear sizes with
redshift and luminosity. We note, that the range of parameter $\beta$
used covers the estimate obtained for kiloparsec-scale structures in
FRII sources by Buchalter et al. (1998; $\beta \approx -0.13\pm 0.06$)
and is close to the estimate obtained earlier for kiloparsec-scale
structures in quasars by Singal (1993; $\beta \approx -0.23\pm 0.12$).
Similarly, our range of the parameter $n$ is close to the estimates
obtained in both papers for kiloparsec-scale structures (Singal 1993,
Buchalter et  al. 1998). However, one must keep in mind that closeness
of these values for kiloparsec-scale structures in double radio sources
and in our parsec-scale structures could be superficial since the radio
emission on these scales, differed by several orders of magnitude, is
governed by different physical processes.

As a test of our method of using median values for binned data, we
repeated the same procedure for the sub-sample of 145 sources in which
the upper limits of angular size (shown in Table~\ref{TSou1} with the
sign ``$<$'' in column 8) are replaced with an arbitrary value of 0.1
mas.  The difference between estimates of $lh$ and $q_{0}$ for this
test case and values presented in Table~\ref{Tregr} does not exceed
4.1\% within the range of $\beta$ and $n$ shown in Table~\ref{Tregr}.
We therefore conclude that the use of median values is justified.

Values of $q_{0}$ shown in Table~\ref{Tregr} should be treated with
caution due to the deficiencies of the sample and the method described
above. For the simple case with no dependence of the source linear size
on the source luminosity and redshift (``true'' standard rod, $\beta =
0$ and $n = 0$) $q_{0} = 0.21 \pm 0.30$. This result does not
contradict to the estimate of $q_{\circ}$ in $\Lambda = 0$,
''no-evolution'' ($\beta = n = 0$) by Buchalter et al. (1998).
Solutions, which allow evolution of source size with redshift ($n \neq
0$) or a dependence on luminosity ($\beta \neq 0$), favor values of
$q_{0} \la 0.5$ for $\beta + n \ga -0.15$.

\section{ Summary }

The 5 GHz VLBI data are consistent with standard FRW cosmologies with
$0 \la q_{\circ} \la 0.5$ and $\Lambda = 0$ without the need to
introduce evolution of the population or to appeal to selection effects
caused by a possible ``luminosity -- linear size'' dependence. This
conclusion is based on the ``angular size -- redshift'' test using an
inhomogeneous sample of 330 VLBI images, with the 1.5 mas nominal
angular resolution and the dynamic range at least 100.  A two-parameter
regression model applied for a plausible range of dependence of linear
size on luminosity and redshift is used to separate the ``$\beta -
n$''parameter space and gives a deceleration parameter somewhat lower
than the critical, $q_{\circ} \le 0.5$, for $\beta + n \ga -0.15$. Such
an approach might be useful to further restrict the deceleration
parameter using a better understanding of physics of the compact radio
structures, represented by parameters $\beta$ and $n$.

We also find a dependence of angular size with spectral index which, if
not considered, increases the dispersion in linear size.  Elimination
of extreme values of spectral indices with $\alpha < -0.38$ and $\alpha
> 0.18$ better defines compact sources as standard rods.

In view of the size and selection bias in the currently available
sample, we have chosen not to consider more general models with
$\Lambda \neq 0$. However, we present our full data set for those may
wish to use these data to further investigate constraints on the
cosmological parameters.

\begin{acknowledgements}

We are grateful to the referee for helpful comments. LIG acknowledges
partial support from the European Union under contract No.~
CHGECT~920011, the Netherlands Organization for Scientific Research
(NWO) programme on the Early Universe, and the European Commission, TMR
Programme, Research Network Contract ERBFMRXCT~96--0034 ``CERES''. SF
acknowledges financial support received from the NWO and the Hungarian
Space Office, and hospitality of JIVE and NFRA during his fellowship in
Dwingeloo. KIK acknowledges the hospitality of JIVE during two
visits. We acknowledge the use of the NASA/IPAC Extragalactic Database
(NED), which is operated by the Jet Propulsion Laboratory, California
Institute of Technology, under contract with the National Aeronautics
and Space Administration. The National Radio Astronomy Observatory is
operated by Associated Universities, Inc.  under a Cooperative
Agreement with the National Science Foundation.

\end{acknowledgements}

%
   \begin{table*}
      \caption{The sample of sources for $\theta-z$ studies. This table
      is available in electronic form at the CDS via anonymous ftp to
      {\tt cdsarc.u-strasbg.fr (130.79.128.5)} or via {\tt
      http://cdsweb.u-strasbg.fr/Abstract.html}}
      \label{TSou1}
\begin{tabular}{clrcrrrrcllll}                     \hline
      \noalign{\smallskip}

                Source         &    
\,\,\,\,        Name           &    
$z$\,\,\,                      &    
ID$^{\rm a}$                   &    
$S_6$\,\,                      &    
$S_{20}$\,                     &    
$\alpha$\,\,\,                 &    
$\theta^{\rm b}$\,\,           &    
Type$^{\rm c}$                 &    
\,              Ref            &    
\,              Ref$^{\rm d}$  &    
\,              Ref$^{\rm e}$  &    
\,              Ref            \\   
                               &    
                               &    
                               &    
                               &    
[Jy]\,                         &    
[Jy]\,                         &    
                               &    
[mas]                          &    
                               &    
\,\,\, $z$                     &    
\,\, $S_6$                     &    
\,   $S_{20}$                  &    
VLBI                           \\   
                        1      &    
\,\,\,\,\,\,\,\,\,\,    2      &    
3\,\,\,\,                      &    
4                              &    
5\,\,\,\,                      &    
6\,\,\,\,                      &    
7\,\,\,\,                      &    
8\,\,\,\,\,                    &    
9                              &    
\,\,                   10      &    
\,\,                   11      &    
\,\,                   12      &    
\,\,                   13      \\   
      \noalign{\smallskip}                                     
      \hline                                                   
      \noalign{\smallskip}                                     
0003+380   &           & 0.229 & Q  & 0.57 &  0.60 & $-$0.04 & $<$2.0 & J  & VV96 &      &      & HB95 \\
0004+139   &           & 3.20  & Q  & 0.12 &  0.19 & $-$0.23 &    5.1 & C  & MH98 &      &      & PF98 \\
0010+405   & 4C 40.01  & 0.255 & G  & 1.04 &  1.80 & $-$0.44 & $<$2.0 & L  & SK93 &      &      & XR95 \\
0014+813   &           & 3.387 & Q  & 0.55 &  [f]  & $-$0.16 &    4.9 & C  & VV96 & KP81 & KP81 & TV94 \\
0016+731   &           & 1.781 & Q  & 1.58 &  0.86 &    0.49 & $<$2.0 & S  & VV96 &      &      & PR88 \\[0.15cm]
0022+390   &           & 1.946 & Q  & 0.71 &  0.76 & $-$0.05 &    9.8 & C  & VV96 &      &      & XR95 \\
0035+367   &           & 0.366 & Q  & 0.48 &  0.88 & $-$0.49 &    3.4 & C  & VV96 &      &      & TV96 \\
0035+413   &           & 1.353 & Q  & 1.14 &  0.44 &    0.77 & $<$0.9 & S  & VV96 &      &      & HB95 \\
0046+063   &           & 3.580 & Q  & 0.21 &  0.22 & $-$0.02 &    4.0 & C  & MH98 &      &      & PF98 \\
0055+301   & NGC 315   & 0.016 & G  & 0.91 &  1.59 & $-$0.45 &   10.2 & C  & DV91 &      &      & VG93 \\[0.15cm]
0106+013   &           & 2.107 & Q  & 4.18 &  4.04 &    0.03 &   10.6 & C  & VV96 &      &      & SW97 \\
0108+388   &           & 0.669 & G  & 1.34 &  0.41 &    0.95 &    5.1 & C  & HR92 &      &      & PR88 \\
0110+495   &           & 0.395 & Q  & 0.72 &  0.49 &    0.31 &    8.1 & C  & VV96 &      &      & HB95 \\
0133+476   & DA 55     & 0.859 & Q  & 2.02 &  1.40 &    0.30 &    2.9 & C  & VV96 &      &      & CW93 \\
0145+386   &           & 1.440 & Q  & 0.36 &  0.29 &    0.17 &    3.2 & C  & HB93 &      &      & HB95 \\[0.15cm]
0151+474   &           & 1.026 & Q  & 0.50 &  0.35 &    0.29 & $<$1.7 & L  & VV96 &      &      & HB95 \\
0153+744   &           & 2.338 & Q  & 1.59 &  2.09 & $-$0.22 &   10.6 & C  & VV96 &      &      & WS88 \\
0201+365   &           & 2.912 & Q  & 0.36 &  0.59 & $-$0.40 &    6.8 & C  & VV96 &      &      & HB95 \\
0205+722   &           & 0.895 & G  & 0.53 &  0.84 & $-$0.37 &    3.3 & C  & VT96 &      &      & TV94 \\
0208$-$512 &           & 1.003 & Q  & 3.30 &  [g]  & $-$0.12 &    1.7 & C  & VV96 & GV94 & WO90 & TE96 \\[0.15cm]
0212+735   &           & 2.367 & Q  & 2.27 &  2.62 & $-$0.12 &   13.1 & C  & VV96 &      &      & PR88 \\
0219+428   &           & 0.444 & B  & 0.99 &  [h]  & $-$0.66 &    5.3 & C  & VV96 &      & FG85 & TV96 \\
0227+403   &           & 1.019 & Q  & 0.41 &  0.44 & $-$0.06 &    3.6 & C  & HB97 &      &      & HB95 \\
0234+285   & CTD 20    & 1.207 & Q  & 2.79 &  2.33 &    0.15 &    3.5 & C  & VV96 &      &      & WC92 \\
0235+164   &           & 0.940 & B  & 1.94 &  2.36 & $-$0.16 &    2.1 & C  & VV96 &      &      & CB96 \\[0.15cm]
0243+181   &           & 3.59  & Q  & 0.22 &  0.19 &    0.12 &    4.6 & C  & MH98 &      &      & PF98 \\
0248+430   &           & 1.310 & Q  & 1.37 &  0.83 &    0.40 &   11.4 & C  & VV96 &      &      & XR95 \\
0249+383   &           & 1.122 & Q  & 0.45 &  0.78 & $-$0.44 &    6.3 & C  & HB97 &      &      & HB95 \\
0251+393   &           & 0.291 & Q  & 0.35 &  0.30 &    0.12 &    1.6 & C  & VV96 &      &      & HB95 \\
0256+424   &           & 0.867 & Q  & 0.39 &  0.62 & $-$0.37 &   20.9 & C  & VV96 &      &      & HB95 \\[0.15cm]
0307+380   &           & 0.816 & Q  & 0.63 &  0.12 &    1.33 & $<$1.0 & J  & VV96 &      & VL96 & HB95 \\
0309+411   & NRAO 128  & 0.134 & G  & 0.53 &  0.47 &    0.10 &    3.9 & C  & MB96 &      &      & HB95 \\
0316+413   & 3C 84     & 0.018 & G  &46.89 & 21.20 &    0.64 &   10.2 & C  & SH92 &      &      & RB95 \\
0332$-$403 &           & 1.445 & Q  & 2.60 &  1.92 &    0.23 &    2.0 & C  & VV96 & WO90 & WO90 & SW98 \\
0333+321   & NRAO 140  & 1.259 & Q  & 1.95 &  3.08 & $-$0.37 &    9.8 & C  & VV96 &      &      & Ma88 \\[0.15cm]
0336$-$019 & CTA 26    & 0.852 & Q  & 2.58 &  2.25 &    0.08 &    1.3 & C  & VV96 & KN81 &      & WC92 \\
0400+258   &           & 2.109 & Q  & 0.99 &  1.48 & $-$0.32 &    4.4 & C  & VV96 &      &      & Ru88 \\
0403$-$132 &           & 0.571 & Q  & 3.24 &  4.00 & $-$0.17 & $<$1.4 & L  & VV96 & WO90 & WO90 & SW98 \\
0410+110   & 3C 109    & 0.306 & G  & 1.39 &  3.93 & $-$0.84 &    3.5 & C  & HB91 &      &      & GF94 \\
0415+379   & 3C 111    & 0.049 & G  & 5.17 & 13.53 & $-$0.77 &    9.5 & C  & HB91 &      &      & PA90 \\[0.15cm]
0420$-$014 &           & 0.915 & Q  & 4.15 &  2.24 &    0.50 &    1.7 & C  & VV96 & WO90 & WO90 & HV98 \\
0430+052   & 3C 120    & 0.033 & G  & 4.20 &  3.85 &    0.07 &   11.8 & C  & MH88 &      &      & WB87 \\
0444+634   &           & 0.781 & Q  & 0.52 &  0.37 &    0.27 &    5.3 & C  & VV96 &      &      & TV94 \\
0454$-$234 &           & 1.003 & Q  & 2.00 &  [x]  &    0.21 &    2.4 & C  & VV96 & WO90 & WO90 & SW98 \\
0454+844   &           & 0.112 & B  & 1.40 &  [i]  &    0.38 &    1.3 & C  & VV96 & KP81 & KP81 & WS88 \\[0.15cm]
0458$-$020 &           & 2.286 & Q  & 2.19 &  2.20 &    0.00 &    4.6 & C  & VV96 & WO90 & WO90 & WC92 \\
0521$-$365 &           & 0.055 & G  & 9.23 & 16.30 & $-$0.46 &    3.9 & C  & Ke85 & WO90 & WO90 & TE96 \\
0537$-$441 &           & 0.896 & B  & 3.80 &  2.70 &    0.28 &    3.9 & C  & VV96 & WO90 & WO90 & TE96 \\
0537+531   &           & 1.275 & Q  & 0.67 &  0.66 &    0.01 &    2.3 & C  & VV96 &      &      & TV94 \\
0546+726   &           & 1.555 & Q  & 0.39 &  0.49 & $-$0.18 &    4.4 & C  & HB97 &      &      & TV94 \\
\noalign{\smallskip}
\hline
\end{tabular}

   \end{table*}

\setcounter{table}{0}
%
   \begin{table*}
      \caption{{\it continued}}
      \label{TSou2}
\begin{tabular}{clrcrrrrcllll}                     \hline
      \noalign{\smallskip}
                        1      &    
\,\,\,\,\,\,\,\,\,\,    2      &    
3\,\,\,\,                      &    
4                              &    
5\,\,\,\,                      &    
6\,\,\,\,                      &    
7\,\,\,\,                      &    
8\,\,\,\,\,                    &    
9                              &    
\,\,                   10      &    
\,\,                   11      &    
\,\,                   12      &    
\,\,                   13      \\   
      \noalign{\smallskip}
      \hline
      \noalign{\smallskip}
0552+398   & DA 193    & 2.365 & Q  & 5.52 &  1.75 &    0.92 & $<$1.0 & J  & VV96 &      &      & WC92 \\
0554+580   &           & 0.904 & Q  & 0.85 &  0.37 &    0.67 &    3.4 & C  & HB97 &      &      & TV94 \\
0600+442   &           & 1.136 & Q  & 0.71 &  1.21 & $-$0.43 &   10.1 & C  & VV96 &      &      & HB95 \\
0601+579   &           & 1.840 & Q  & 0.16 &  0.14 &    0.11 & $<$1.3 & J  & Sn97 &      &      & Sn97 \\
0609+607   &           & 2.702 & Q  & 1.07 &  1.06 &    0.01 &    4.9 & C  & HB97 &      &      & TV94 \\[0.15cm]
0615+820   &           & 0.71  & Q  & 1.00 &  0.78 &    0.20 & $<$1.0 & L  & VV96 & KP81 &      & XR95 \\
0620+389   &           & 3.469 & Q  & 0.84 &  1.22 & $-$0.30 &    6.5 & C  & VV96 &      &      & XR95 \\
0627+532   &           & 2.204 & Q  & 0.45 &  0.81 & $-$0.47 &   10.2 & C  & HB97 &      &      & HB95 \\
0633+734   &           & 1.850 & Q  & 0.71 &  1.10 & $-$0.35 &    5.7 & C  & HB97 &      &      & TV94 \\
0636+680   &           & 3.177 & Q  & 0.49 &  0.13 &    1.07 & $<$1.5 & L  & VV96 &      &      & TV94 \\[0.15cm]
0641+393   &           & 1.266 & Q  & 0.44 &  0.37 &    0.14 &    3.6 & C  & HB97 &      &      & HB95 \\
0642+449   &           & 3.408 & Q  & 1.22 &  0.60 &    0.57 &    3.1 & C  & VV96 &      &      & XR95 \\
0646+600   &           & 0.455 & Q  & 0.94 &  0.44 &    0.61 &    3.0 & C  & VV96 &      &      & XR95 \\
0650+371   &           & 1.982 & Q  & 0.87 &  0.59 &    0.31 & $<$0.8 & J  & VV96 &      &      & XR95 \\
0650+453   &           & 0.933 & Q  & 0.47 &  0.59 & $-$0.18 & $<$1.0 & J  & HB97 &      &      & HB95 \\[0.15cm]
0651+410   &           & 0.022 & G  & 0.43 &  0.30 &    0.29 & $<$2.0 & L  & MH96 &      &      & HB95 \\
0707+476   &           & 1.292 & Q  & 0.98 &  0.98 &    0.00 &    3.5 & C  & VV96 &      &      & XR95 \\
0710+439   &           & 0.518 & Q  & 1.61 &  1.83 & $-$0.10 &   24.0 & C  & VV96 &      &      & CP92 \\
0711+356   &           & 1.626 & Q  & 0.82 &  1.43 & $-$0.45 &    5.2 & C  & VV96 &      &      & PR88 \\
0714+457   &           & 0.940 & Q  & 0.47 &  0.41 &    0.11 &    3.9 & C  & VV96 &      &      & HB95 \\[0.15cm]
0724+571   &           & 0.426 & Q  & 0.39 &  0.41 & $-$0.04 &    4.3 & C  & HB97 &      &      & TV94 \\
0727+409   &           & 2.501 & Q  & 0.47 &  0.41 &    0.11 &    4.0 & C  & VV96 &      &      & HB95 \\
0730+504   &           & 0.720 & Q  & 0.99 &  0.39 &    0.75 &    2.5 & C  & HB97 &      &      & TV94 \\
0731+479   &           & 0.782 & Q  & 0.51 &  0.43 &    0.14 &    3.6 & C  & VV96 &      &      & HB95 \\
0740+828   &           & 1.991 & Q  & 0.93 &  1.82 & $-$0.54 &    8.9 & C  & VV96 & KP81 &      & XR95 \\[0.15cm]
0743+744   &           & 1.629 & Q  & 0.48 &  0.35 &    0.25 &    2.2 & C  & VV96 &      &      & HB95 \\
0746+483   &           & 1.951 & Q  & 0.90 &  0.66 &    0.25 &    3.0 & C  & VV96 &      &      & XR95 \\
0754+100   &           & 0.66  & B  & 0.90 &  1.05 & $-$0.12 &    1.5 & C  & VV96 &      &      & GC92 \\
0755+379   & NGC 2484  & 0.043 & G  & 1.11 &  2.58 & $-$0.68 &    5.6 & C  & DV91 &      &      & XR95 \\
0758+595   &           & 1.977 & Q  & 0.18 &  0.14 &    0.20 &    2.0 & C  & Sn97 &      &      & Sn97 \\[0.15cm]
0803+452   &           & 2.102 & Q  & 0.38 &  0.39 & $-$0.02 &    2.1 & C  & HB97 &      &      & HB95 \\
0804+499   &           & 1.433 & Q  & 1.32 &  0.89 &    0.32 & $<$1.2 & S  & VV96 &      &      & PR88 \\
0805+410   &           & 1.420 & Q  & 0.69 &  0.36 &    0.52 &    3.0 & C  & VV96 &      &      & XR95 \\
0806+573   &           & 0.611 & Q  & 0.41 &  0.43 & $-$0.04 &   25.1 & C  & HB97 &      &      & TV94 \\
0812+367   &           & 1.025 & Q  & 0.99 &  1.02 & $-$0.02 &   10.6 & C  & VV96 &      &      & XR95 \\[0.15cm]
0820+560   &           & 1.417 & Q  & 1.16 &  1.27 & $-$0.07 &    2.7 & C  & VV96 &      &      & XR95 \\
0821+394   &           & 1.216 & Q  & 1.03 &  1.38 & $-$0.24 &    4.3 & C  & VV96 &      &      & XR95 \\
0821+621   &           & 0.542 & Q  & 0.62 &  0.65 & $-$0.04 &   32.6 & C  & VV96 &      &      & TV94 \\
0824+355   &           & 2.249 & Q  & 0.75 &  0.87 & $-$0.12 & $<$1.0 & J  & VV96 &      &      & HB95 \\
0826+707   &           & 2.003 & Q  & 0.11 &  0.07 &    0.36 & $<$1.3 & J  & Sn97 &      & Sn97 & Sn97 \\[0.15cm]
0828+493   &           & 0.548 & B  & 0.37 &  1.00 & $-$0.80 &    2.2 & C  & VV96 &      &      & XR95 \\
0830+102   &           & 3.750 & Q  & 0.13 &  0.15 & $-$0.11 &   14.2 & C  & OW95 &      &      & PF98 \\
0830+425   &           & 0.253 & Q  & 0.39 &  0.24 &    0.39 &    4.7 & C  & HB97 &      & WB97 & HB95 \\
0831+557   & 4C 55.16  & 0.242 & G  & 5.74 &  7.74 & $-$0.24 &    5.3 & C  & AA92 &      &      & PR88 \\
0833+416   &           & 1.298 & Q  & 0.39 &  0.43 & $-$0.08 &    4.0 & C  & HB97 &      &      & HB95 \\[0.15cm]
0833+585   &           & 2.101 & Q  & 0.72 &  0.60 &    0.15 &    1.2 & C  & VV96 &      &      & XR95 \\
0836+290   & 4C 29.30  & 0.079 & G  & 0.27 &  0.88 & $-$0.95 &    6.3 & C  & OL95 &      &      & VC95 \\
0836+710   &           & 2.172 & Q  & 2.34 &  4.24 & $-$0.48 &    8.7 & C  & VV96 &      &      & WS88 \\
0850+581   & 4C 58.17  & 1.322 & Q  & 1.18 &  1.42 & $-$0.15 &    6.0 & C  & VV96 &      &      & HS92 \\
0851+202   & OJ287     & 0.306 & Q  & 2.91 &  2.28 &    0.20 &    3.0 & C  & VV96 &      &      & GW89 \\
\noalign{\smallskip}                       
\hline                                     
\end{tabular}                              
                                           
   \end{table*}

\setcounter{table}{0}
%
   \begin{table*}                          
      \caption{{\it continued}}                           
      \label{TSou3}                        
\begin{tabular}{clrcrrrrcllll}                    \hline
      \noalign{\smallskip}                 
                        1      &    
\,\,\,\,\,\,\,\,\,\,    2      &    
3\,\,\,\,                      &    
4                              &    
5\,\,\,\,                      &    
6\,\,\,\,                      &    
7\,\,\,\,                      &    
8\,\,\,\,\,                    &    
9                              &    
\,\,                   10      &    
\,\,                   11      &    
\,\,                   12      &    
\,\,                   13      \\   
     \noalign{\smallskip}                 
      \hline                               
      \noalign{\smallskip}                 
0859+681   &           & 1.499 & Q  & 0.66 &  0.59 &    0.09 &    4.9 & C  & VV96 &      &      & TV94 \\
0900+520   &           & 1.537 & Q  & 0.37 &  0.32 &    0.12 &    1.4 & C  & HB97 &      &      & TV94 \\
0902+490   &           & 2.690 & Q  & 0.55 &  0.64 & $-$0.12 & $<$1.0 & S  & VV96 &      &      & HB95 \\
0906+041   &           & 3.20  & Q  & 0.13 &  0.21 & $-$0.38 &    7.1 & C  & BS95 &      &      & PF98 \\
0906+430   & 3C 216    & 0.668 & Q  & 1.61 &  4.27 & $-$0.79 &    3.5 & C  & VV96 &      &      & VP93 \\[0.15cm]
0913+391   &           & 1.269 & Q  & 0.55 &  1.06 & $-$0.53 &    3.4 & C  & VV96 &      &      & HB95 \\
0917+449   &           & 2.180 & Q  & 1.09 &  0.78 &    0.27 &    2.8 & C  & VV96 &      &      & XR95 \\
0917+624   &           & 1.446 & Q  & 1.23 &  1.23 &    0.00 &    5.1 & C  & VV96 &      &      & SQ96 \\
0923+392   & 4C 39.25  & 0.698 & Q  & 6.91 &  2.72 &    0.75 &    2.0 & C  & VV96 &      &      & PR88 \\
0929+533   &           & 0.595 & Q  & 0.39 &  0.54 & $-$0.26 &    5.6 & C  & VV96 &      &      & TV94 \\[0.15cm]
0930+493   &           & 2.582 & Q  & 0.53 &  0.73 & $-$0.26 &    2.3 & C  & HB97 &      &      & HB95 \\
0933+503   &           & 0.276 & G  & 0.32 &  0.14 &    0.67 & $<$1.0 & J  & HB97 &      & VL96 & HB95 \\
0938+119   &           & 3.191 & Q  & 0.12 &  0.29 & $-$0.70 &    2.8 & C  & VV96 &      &      & PF98 \\
0941+522   &           & 0.565 & Q  & 0.39 &  0.85 & $-$0.63 &    6.8 & C  & VV96 &      &      & HB95 \\
0945+408   & 4C 40.24  & 1.252 & Q  & 1.80 &  1.49 &    0.15 &    8.3 & C  & VV96 &      &      & PR88 \\[0.15cm]
0949+354   &           & 1.875 & Q  & 0.37 &  0.34 &    0.07 &    8.5 & C  & VV96 &      &      & HB95 \\
0954+658   &           & 0.367 & Q  & 1.13 &  0.65 &    0.45 &    1.8 & C  & VV96 &      &      & GM94 \\
0955+476   &           & 1.873 & Q  & 1.01 &  0.69 &    0.31 & $<$1.8 & J  & VV96 &      &      & XR95 \\
1003+830   &           & 0.322 & G  & 0.72 &  0.60 &    0.15 &    6.0 & C  & XL94 & KP81 &      & XR95 \\
1010+350   &           & 1.414 & Q  & 0.63 &  0.42 &    0.33 &    8.1 & C  & VV96 &      &      & HB95 \\[0.15cm]
1020+400   &           & 1.254 & Q  & 0.79 &  1.16 & $-$0.31 &    3.4 & C  & VV96 &      &      & XR95 \\
1030+398   &           & 1.095 & Q  & 0.65 &  0.38 &    0.43 &    2.0 & C  & VV96 &      &      & HB95 \\
1030+415   &           & 1.120 & Q  & 0.44 &  0.77 & $-$0.45 &    4.2 & C  & VV96 &      &      & XR95 \\
1030+611   &           & 0.336 & G  & 0.53 &  0.77 & $-$0.30 &    3.8 & C  & VV96 &      &      & TV94 \\
1034$-$293 &           & 0.312 & B  & 1.51 &  [j]  &    0.21 & $<$1.3 & L  & VV96 & WO90 & WO90 & HV98 \\[0.15cm]
1038+528   &           & 0.677 & Q  & 0.70 &  0.71 & $-$0.01 & $<$1.5 & J  & VV96 &      &      & HB95 \\
1039+811   &           & 1.256 & Q  & 1.14 &  0.73 &    0.36 &    2.3 & C  & VV96 & KP81 &      & XR95 \\
1041+536   &           & 1.897 & Q  & 0.44 &  0.54 & $-$0.16 &    3.3 & C  & HB97 &      &      & HB95 \\
1044+719   &           & 1.15  & Q  & 1.90 &  0.62 &    0.90 & $<$1.5 & L  & VV96 &      &      & XR95 \\
1053+704   &           & 2.492 & Q  & 0.54 &  0.61 & $-$0.10 &    1.9 & C  & VV96 &      &      & XR95 \\[0.15cm]
1053+815   &           & 0.706 & G  & 0.77 &  0.34 &    0.66 & $<$1.5 & J  & XL94 & KP81 &      & XR95 \\
1055+201   &           & 1.11  & Q  & 1.51 &  2.31 & $-$0.34 & $<$2.0 & L  & VV96 &      &      & HS92 \\
1058+726   &           & 1.46  & Q  & 0.86 &  1.45 & $-$0.42 &   19.2 & C  & VV96 &      &      & XR95 \\
1058+629   &           & 0.664 & Q  & 0.69 &  0.60 &    0.11 &    1.4 & C  & HB97 &      &      & XR95 \\
1101+384   & Mrk421    & 0.031 & B  & 0.72 &  0.84 & $-$0.12 &   12.8 & C  & VV96 &      &      & XR95 \\[0.15cm]
1104$-$445 &           & 1.598 & Q  & 2.03 &  1.92 &    0.04 &    2.7 & C  & VV96 & WO90 & WO90 & SW97 \\    
1105+437   &           & 1.226 & Q  & 0.34 &  0.27 &    0.19 &    1.6 & C  & HB97 &      &      & HB95 \\
1124+571   &           & 2.890 & Q  & 0.45 &  0.78 & $-$0.44 &    2.0 & C  & VV96 &      &      & TA94 \\
1127$-$145 &           & 1.187 & Q  & 5.46 &  6.40 & $-$0.13 &   16.1 & C  & VV96 & WO90 & WO90 & WC92 \\
1128+385   &           & 1.733 & Q  & 0.77 &  0.93 & $-$0.15 & $<$1.0 & J  & VV96 &      &      & XR95 \\[0.15cm]
1143+590   &           & 1.982 & Q  & 0.58 &  0.28 &    0.59 & $<$1.1 & J  & HB97 &      &      & TV94 \\
1144+352   &           & 0.063 & G  & 0.67 &  0.70 & $-$0.04 &   21.6 & C  & MB96 &      &      & HB95 \\
1144+542   &           & 2.201 & Q  & 0.52 &  0.41 &    0.19 &    2.5 & C  & VV96 &      &      & XR95 \\
1146+531   &           & 1.629 & Q  & 0.29 &  0.20 &    0.30 &    1.5 & C  & VV96 &      & VL96 & HB95 \\
1146+596   & NGC 3894  & 0.011 & G  & 0.57 &  0.41 &    0.27 &   19.9 & C  & DV91 &      &      & TA94 \\[0.15cm]
1150+497   & 4C 49.22  & 0.334 & Q  & 0.72 &  1.43 & $-$0.55 & $<$1.8 & J  & VV96 &      &      & XR95 \\
1150+812   &           & 1.25  & Q  & 1.18 &  1.38 & $-$0.13 &    2.8 & C  & VV96 & KP81 &      & XR95 \\
1151+408   &           & 0.916 & Q  & 0.37 &  0.70 & $-$0.51 &    1.7 & C  & HB97 &      &      & HB95 \\
1155+486   &           & 2.028 & Q  & 0.55 &  0.48 &    0.11 &    2.2 & C  & HB97 &      &      & HB95 \\
1156+295   &           & 0.729 & Q  & 1.46 &  1.75 & $-$0.15 &    3.6 & C  & VV96 &      &      & MM90 \\
\noalign{\smallskip}                       
\hline                                     
\end{tabular}                              
                                           
   \end{table*}

\setcounter{table}{0}
%
   \begin{table*}                          
      \caption{{\it continued}}                           
      \label{TSou4}                        
\begin{tabular}{clrcrrrrcllll}                    \hline
      \noalign{\smallskip}                 
                        1      &    
\,\,\,\,\,\,\,\,\,\,    2      &    
3\,\,\,\,                      &    
4                              &    
5\,\,\,\,                      &    
6\,\,\,\,                      &    
7\,\,\,\,                      &    
8\,\,\,\,\,                    &    
9                              &    
\,\,                   10      &    
\,\,                   11      &    
\,\,                   12      &    
\,\,                   13      \\   
      \noalign{\smallskip}                 
      \hline                               
      \noalign{\smallskip}                 
1213+350   &           & 0.857 & Q  & 1.12 &  1.73 & $-$0.35 &   36.8 & C  & VV96 &      &      & XR95 \\
1214+588   &           & 2.547 & Q  & 0.31 &  0.42 & $-$0.24 &    2.0 & C  & HB97 &      &      & TV94 \\
1216+487   &           & 1.076 & Q  & 0.64 &  0.86 & $-$0.24 &    5.8 & C  & VV96 &      &      & XR95 \\
1222+216   &           & 0.435 & Q  & 1.15 &  1.97 & $-$0.43 &    3.4 & C  & VV96 &      &      & HO92 \\
1223+395   &           & 0.623 & Q  & 0.51 &  0.54 & $-$0.05 &   17.4 & C  & VV96 &      &      & HB95 \\[0.15cm]
1225+368   &           & 1.975 & Q  & 0.79 &  2.14 & $-$0.80 &   31.3 & C  & VV96 &      &      & XR95 \\
1226+023   & 3C 273    & 0.158 & Q  &43.57 & 50.10 & $-$0.11 &   26.1 & C  & VV96 &      &      & ZB88 \\
1226+373   &           & 1.515 & Q  & 0.86 &  0.19 &    1.22 & $<$0.9 & J  & HB97 &      &      & HB95 \\
1239+376   &           & 3.818 & Q  & 0.37 &  0.54 & $-$0.30 & $<$2.0 & L  & VT96 &      &      & HB95 \\
1240+381   &           & 1.316 & Q  & 0.76 &  0.36 &    0.60 & $<$1.0 & J  & VV96 &      &      & HB95 \\[0.15cm]
1244$-$255 &           & 0.638 & Q  & 1.55 &  [y]  &    0.24 & $<$1.6 & L  & VV96 & WO90 & WO90 & SW98 \\
1253$-$055 & 3C 279    & 0.538 & Q  &13.00 & 11.60 &    0.09 &    3.2 & C  & VV96 & WO90 & WO90 & UC89 \\
1254+571   &           & 0.042 & G  & 0.42 &  0.29 &    0.30 & $<$1.5 & J  & DS93 &      &      & TV94 \\
1258+507   &           & 1.561 & Q  & 0.44 &  0.52 & $-$0.13 & $<$1.5 & J  & VV96 &      &      & HB95 \\
1305+804   &           & 1.183 & Q  & 0.38 &  0.86 & $-$0.66 &   13.3 & C  & VT96 & KP81 &      & TV96 \\[0.15cm]
1307+562   &           & 1.629 & Q  & 0.42 &  0.29 &    0.30 &    1.9 & C  & HB97 &      &      & TV94 \\
1308+326   &           & 0.997 & Q  & 1.45 &  1.61 & $-$0.08 &    3.8 & C  & VV96 &      &      & GC92 \\
1309+555   &           & 0.926 & Q  & 0.68 &  0.21 &    0.95 & $<$1.4 & J  & HB97 &      &      & TV94 \\
1311+552   &           & 0.613 & Q  & 0.55 &  1.17 & $-$0.61 &   38.4 & C  & VT96 &      &      & TV94 \\
1317+520   &           & 1.055 & Q  & 0.61 &  1.29 & $-$0.60 &    8.7 & C  & VV96 &      &      & HS92 \\[0.15cm]
1321+410   &           & 0.496 & Q  & 0.41 &  0.36 &    0.10 &    5.4 & C  & VT96 &      &      & HB95 \\
1323+800   &           & 1.970 & Q  & 0.46 &  [k]  &    0.22 &    3.5 & C  & VV96 & KP81 & KP81 & TV94 \\
1325+436   &           & 2.073 & Q  & 0.58 &  0.70 & $-$0.15 & $<$1.0 & J  & VV96 &      &      & HB95 \\
1333+459   &           & 2.450 & Q  & 0.65 &  0.31 &    0.60 & $<$1.0 & J  & VV96 &      &      & XR95 \\
1334$-$127 &           & 0.539 & Q  & 4.30 &  1.90 &    0.66 &    1.6 & C  & VV96 & VL96 & VL96 & HV98 \\[0.15cm]
1335+552   &           & 1.096 & Q  & 0.75 &  0.72 &    0.03 &    2.1 & C  & VV96 &      &      & TV94 \\
1337+637   &           & 2.558 & Q  & 0.42 &  0.50 & $-$0.14 &    7.0 & C  & VV96 &      &      & TV94 \\
1338+381   &           & 3.103 & Q  & 0.26 &  0.33 & $-$0.11 &    3.8 & C  & VV96 &      &      & PF98 \\
1342+662   &           & 0.766 & Q  & 0.30 &  0.89 & $-$0.88 & $<$1.2 & J  & VV96 &      &      & TV94 \\
1342+663   &           & 1.351 & Q  & 0.55 &  0.89 & $-$0.39 & $<$1.3 & L  & VV96 &      &      & XR95 \\[0.15cm]
1347+539   &           & 0.976 & Q  & 0.64 &  1.15 & $-$0.47 &   12.8 & C  & VV96 &      &      & XR95 \\
1354$-$174 &           & 3.147 & Q  & 0.97 &  1.90 & $-$0.54 & $<$1.9 & J  & VV96 & WO90 & WO90 & FG97 \\
1354+195   &           & 0.719 & Q  & 2.62 &  2.63 &    0.00 &   12.3 & C  & VV96 &      &      & Ru88 \\
1356+478   &           & 0.230 & G  & 0.43 &  0.59 & $-$0.25 &    6.0 & C  & VT96 &      &      & TV96 \\
1402+044   &           & 3.211 & Q  & 1.00 &  0.56 &    0.47 &   14.5 & C  & VV96 &      &      & GK92 \\[0.15cm]
1404+286   & OQ208     & 0.077 & G  & 2.35 &  0.76 &    0.91 &    6.9 & C  & VV96 &      &      & ZB94 \\
1413+135   &           & 0.247 & B  & 0.85 &  1.21 & $-$0.28 &   34.8 & C  & VV96 &      &      & PC96 \\
1413+373   &           & 2.36  & Q  & 0.38 &  0.37 &    0.02 &    4.0 & C  & VV96 &      &      & HB95 \\
1415+463   &           & 1.552 & Q  & 0.80 &  1.01 & $-$0.19 &   10.5 & C  & VV96 &      &      & HB95 \\
1417+385   &           & 1.832 & Q  & 0.65 &  0.71 & $-$0.07 & $<$0.9 & S  & VV96 &      &      & HB95 \\[0.15cm]
1418+546   &           & 0.152 & B  & 1.35 &  1.56 & $-$0.12 &    3.8 & C  & VV96 &      &      & XR95 \\
1421+482   &           & 2.220 & Q  & 0.52 &  0.36 &    0.30 &    2.9 & C  & VV96 &      &      & HB95 \\
1424+366   &           & 1.091 & Q  & 0.44 &  0.19 &    0.68 & $<$0.9 & J  & HB97 &      &      & HB95 \\
1427+543   &           & 2.991 & Q  & 0.72 &  0.90 & $-$0.18 &    9.8 & C  & HB97 &      &      & HB95 \\
1428+422   &           & 4.715 & Q  & 0.34 &  0.31 &    0.07 & $<$1.2 & L  & HM97 &      &      & PF98 \\[0.15cm]
1432+422   &           & 1.240 & Q  & 0.35 &  0.28 &    0.18 & $<$1.6 & J  & VT96 &      &      & TV96 \\
1435+638   &           & 2.062 & Q  & 0.76 &  1.39 & $-$0.49 &    8.9 & C  & VV96 &      &      & XR95 \\
1438+385   &           & 1.775 & Q  & 0.89 &  1.03 & $-$0.12 &    8.8 & C  & VT96 &      &      & XR95 \\
1442+101   &           & 3.535 & Q  & 1.28 &  2.42 & $-$0.51 &   13.5 & C  & VV96 &      &      & UT97 \\
1442+637   &           & 1.380 & Q  & 0.44 &  0.68 & $-$0.35 &    8.7 & C  & VV96 &      &      & TV94 \\
\noalign{\smallskip}                       
\hline                                     
\end{tabular}                              
                                           
   \end{table*}

\setcounter{table}{0}
%
   \begin{table*}                          
      \caption{{\it continued}}                           
      \label{TSou5}                        
\begin{tabular}{clrcrrrrcllll}                    \hline
      \noalign{\smallskip}                 
                        1      &    
\,\,\,\,\,\,\,\,\,\,    2      &    
3\,\,\,\,                      &    
4                              &    
5\,\,\,\,                      &    
6\,\,\,\,                      &    
7\,\,\,\,                      &    
8\,\,\,\,\,                    &    
9                              &    
\,\,                   10      &    
\,\,                   11      &    
\,\,                   12      &    
\,\,                   13      \\   
      \noalign{\smallskip}                 
      \hline                               
      \noalign{\smallskip}                 
1448+762   &           & 0.899 & Q  & 0.68 &   [l] &    0.32 &    1.9 & C  & VV96 & KP81 & KP81 & HB95 \\
1456+375   &           & 0.333 & G  & 0.54 &  0.34 &    0.37 & $<$1.0 & J  & VT96 &      &      & HB95 \\
1458+718   & 3C 309.1  & 0.904 & Q  & 3.57 &  7.68 & $-$0.62 &   49.8 & C  & VV96 &      &      & KW90 \\
1500+045   &           & 3.67  & Q  & 0.18 &  0.12 &    0.32 &    0.7 & C  & VV96 &      &      & PF98 \\
1504$-$166 &           & 0.876 & Q  & 1.96 &  2.70 & $-$0.25 &    1.7 & C  & VV96 & WO90 & WO90 & HV98 \\[0.15cm]
1504+377   &           & 0.674 & G  & 0.97 &  1.19 & $-$0.16 &   10.7 & C  & SK94 &      &      & XR95 \\
1505+428   &           & 0.587 & Q  & 0.41 &  0.44 & $-$0.06 &    5.5 & C  & VV96 &      &      & HB95 \\
1531+722   &           & 0.899 & Q  & 0.44 &  0.66 & $-$0.33 &    2.2 & C  & VV96 &      &      & TV94 \\
1532+016   &           & 1.435 & Q  & 0.79 &  1.20 & $-$0.33 &    1.2 & C  & VV96 & WO90 & WO90 & HV98 \\
1534+501   &           & 1.119 & Q  & 0.37 &  0.23 &    0.38 & $<$1.8 & L  & VV96 &      &      & HB95 \\[0.15cm]
1538+149   &           & 0.605 & B  & 1.21 &  1.45 & $-$0.15 &    4.5 & C  & VV96 &      &      & GC92 \\
1538+593   &           & 3.878 & Q  & 0.08 &  0.05 &    0.38 & $<$1.1 & J  & Sn97 &      & Sn97 & Sn97 \\
1543+480   &           & 1.277 & Q  & 0.44 &  0.67 & $-$0.34 &   34.6 & C  & VV96 &      &      & HB95 \\
1543+517   &           & 1.924 & Q  & 0.59 &  0.49 &    0.15 &    5.1 & C  & HB97 &      &      & HB95 \\
1547+507   &           & 2.169 & Q  & 0.73 &  0.67 &    0.07 &    7.1 & C  & VV96 &      &      & XR95 \\[0.15cm]
1550+582   &           & 1.319 & Q  & 0.35 &  0.23 &    0.34 & $<$1.4 & J  & HB97 &      &      & HB95 \\
1557+031   &           & 3.891 & Q  & 0.41 &  0.48 & $-$0.13 & $<$1.2 & L  & VV96 &      &      & PF98 \\
1602+576   &           & 2.858 & Q  & 0.37 &  0.79 & $-$0.61 &    3.9 & C  & VV96 &      &      & HB95 \\
1614+051   &           & 3.217 & Q  & 0.92 &  0.31 &    0.88 & $<$1.0 & S  & VV96 &      &      & GK92 \\
1619+491   &           & 1.513 & Q  & 0.44 &  0.47 & $-$0.05 &   10.9 & C  & HB97 &      &      & HB95 \\[0.15cm]
1622$-$297 &           & 0.815 & Q  & 1.86 &  2.20 & $-$0.13 &   16.0 & C  & VV96 & WO90 & WO90 & TM98 \\
1622+665   &           & 0.203 & Q  & 0.52 &  0.20 &    0.77 & $<$2.0 & J  & VV96 &      &      & TV96 \\
1623+578   &           & 0.789 & G  & 0.59 &  0.50 &    0.13 &    3.1 & C  & VT96 &      &      & TV96 \\
1624+416   & 4C 41.32  & 2.55  & Q  & 1.25 &  1.68 & $-$0.24 &    5.4 & C  & VV96 &      &      & PR88 \\
1626+396   & 3C 338    & 0.030 & G  & 0.46 &  3.71 & $-$1.68 &   11.5 & C  & DV91 &      &      & FC93 \\[0.15cm]
1633+382   &           & 1.814 & Q  & 3.22 &  1.90 &    0.42 &    1.7 & C  & VV96 &      &      & PR88 \\
1636+473   &           & 0.740 & Q  & 1.24 &  0.95 &    0.21 & $<$1.9 & J  & VV96 &      &      & HB95 \\
1637+826   & NGC 6251  & 0.023 & G  & 0.70 &  0.40 &    0.45 &    5.0 & C  & DV91 & VL96 & VL96 & JU86 \\
1638+398   &           & 1.666 & Q  & 1.12 &  0.66 &    0.43 & $<$0.9 & J  & VV96 &      &      & XR95 \\
1638+540   &           & 1.977 & Q  & 0.35 &  0.31 &    0.10 &    2.7 & C  & HB97 &      &      & HB95 \\[0.15cm]
1641+399   & 3C 345    & 0.594 & Q  & 8.72 &  7.89 &    0.08 &    5.6 & C  & VV96 &      &      & Lo96 \\
1642+690   & 4C 69.21  & 0.751 & Q  & 1.53 &  1.51 &    0.01 &    4.1 & C  & VV96 &      &      & PR88 \\
1645+410   &           & 0.835 & Q  & 0.40 &  0.32 &    0.18 & $<$2.1 & J  & HB97 &      &      & HB95 \\
1645+635   &           & 2.379 & Q  & 0.48 &  0.30 &    0.38 &    7.4 & C  & HB97 &      &      & TV94 \\
1652+398   & Mrk501    & 0.034 & G  & 1.38 &  1.44 & $-$0.03 &    7.8 & C  & DV91 &      &      & PR88 \\[0.15cm]
1656+477   &           & 1.622 & Q  & 1.24 &  0.78 &    0.37 &    5.7 & C  & VV96 &      &      & XR95 \\
1656+571   &           & 1.290 & Q  & 0.76 &  0.81 & $-$0.05 &    4.4 & C  & VV96 &      &      & TV94 \\
1700+685   &           & 0.301 & Q  & 0.38 &  0.30 &    0.19 &    2.0 & C  & HB97 &      &      & TV94 \\
1716+686   &           & 0.777 & Q  & 0.84 &  0.41 &    0.58 &    1.7 & C  & VV96 &      &      & TV94 \\
1719+357   &           & 0.263 & Q  & 0.78 &  0.84 & $-$0.06 &    4.1 & C  & VV96 &      &      & XR95 \\[0.15cm]
1722+401   &           & 1.049 & Q  & 0.52 &  0.55 & $-$0.05 &    4.5 & C  & VT96 &      &      & HB95 \\
1726+455   &           & 0.714 & Q  & 0.94 &  0.43 &    0.63 & $<$0.9 & J  & VV96 &      &      & HB95 \\
1730$-$130 & NRAO 530  & 0.902 & Q  & 4.10 &  5.20 & $-$0.19 &    4.5 & C  & VV96 & WO90 & WO90 & SW97 \\
1732+389   &           & 0.976 & Q  & 0.56 &  0.78 & $-$0.27 & $<$0.9 & J  & VV96 &      &      & XR95 \\
1738+476   &           & 0.316 & B  & 0.82 &  0.83 & $-$0.01 & $<$0.9 & J  & VV96 &      &      & XR95 \\[0.15cm]
1738+499   &           & 1.545 & Q  & 0.43 &  0.57 & $-$0.23 & $<$1.6 & J  & VV96 &      &      & TV94 \\
1739+522   &           & 1.379 & Q  & 1.70 &  1.98 & $-$0.12 & $<$0.7 & J  & VV96 &      &      & PR88 \\
1741$-$038 &           & 1.057 & Q  & 2.30 &  1.17 &    0.53 &    1.6 & L  & VV96 & WO90 & WO90 & SW97 \\
1743+173   &           & 1.702 & Q  & 0.69 &  1.36 & $-$0.55 &    9.0 & C  & VV96 &      &      & Ru88 \\
1745+624   &           & 3.889 & Q  & 0.59 &  0.76 & $-$0.20 &    2.7 & C  & VV96 &      &      & TV94 \\
\noalign{\smallskip}                       
\hline                                     
\end{tabular}                              
                                           
   \end{table*}

\setcounter{table}{0}
%
   \begin{table*}                          
      \caption{{\it continued}}                           
      \label{TSou6}                        
\begin{tabular}{clrcrrrrcllll}                    \hline
      \noalign{\smallskip}                 
                        1      &    
\,\,\,\,\,\,\,\,\,\,    2      &    
3\,\,\,\,                      &    
4                              &    
5\,\,\,\,                      &    
6\,\,\,\,                      &    
7\,\,\,\,                      &    
8\,\,\,\,\,                    &    
9                              &    
\,\,                   10      &    
\,\,                   11      &    
\,\,                   12      &    
\,\,                   13      \\   
      \noalign{\smallskip}                 
      \hline                               
      \noalign{\smallskip}                 
1746+693   &           & 1.886 & Q  & 0.14 &  0.20 & $-$0.28 &    1.7 & C  & Sn97 &      &      & Sn97 \\
1749+096   &           & 0.320 & Q  & 2.46 &  0.61 &    1.12 &    1.7 & C  & VV96 &      &      & WC92 \\
1749+701   &           & 0.770 & Q  & 0.72 &  1.31 & $-$0.48 &    4.7 & C  & VV96 &      &      & GM94 \\
1751+441   &           & 0.871 & Q  & 1.00 &  0.78 &    0.20 &    1.7 & C  & VV96 &      &      & XR95 \\
1755+578   &           & 2.110 & Q  & 0.46 &  0.73 & $-$0.37 &   10.8 & C  & HB97 &      &      & TV94 \\[0.15cm]
1758+388   &           & 2.092 & Q  & 0.74 &  0.51 &    0.30 &    1.4 & C  & VV96 &      &      & XR95 \\
1800+440   &           & 0.663 & Q  & 1.19 &  0.88 &    0.24 & $<$0.9 & J  & VV96 &      &      & XR95 \\
1803+784   &           & 0.684 & Q  & 2.63 &  1.87 &    0.27 &    1.6 & C  & VV96 & KP81 &      & CW93 \\
1806+456   &           & 0.830 & Q  & 0.35 &  0.15 &    0.68 & $<$1.7 & J  & VV96 &      &      & TV94 \\
1807+698   & 3C 371    & 0.051 & Q  & 2.12 &  2.26 & $-$0.05 &    4.0 & C  & VV96 &      &      & CW93 \\[0.15cm]
1811+430   &           & 1.090 & Q  & 0.51 &  0.97 & $-$0.52 &   10.9 & C  & VV96 &      &      & TV94 \\
1812+412   &           & 1.564 & Q  & 0.52 &  0.64 & $-$0.17 &   10.1 & C  & HB97 &      &      & HB95 \\
1818+356   &           & 0.971 & Q  & 0.57 &  0.99 & $-$0.44 & $<$1.6 & J  & VT96 &      &      & TV96 \\
1823+568   &           & 0.664 & B  & 1.26 &  1.48 & $-$0.13 &    6.1 & C  & VV96 &      &      & GM94 \\
1826+796   &           & 0.224 & Q  & 0.58 &  [m]  &    0.40 &   15.7 & C  & HB97 & KP81 & KP81 & TV94 \\[0.15cm]
1828+487   & 3C 380    & 0.692 & Q  & 5.52 & 14.65 & $-$0.79 &   23.4 & C  & VV96 &      &      & PW93 \\
1830+285   &           & 0.594 & Q  & 0.98 &  1.81 & $-$0.49 &    1.7 & C  & VV96 &      &      & HS92 \\
1834+612   &           & 2.274 & Q  & 0.57 &  0.45 &    0.19 &    3.0 & C  & HB97 &      &      & TV94 \\
1839+389   &           & 3.095 & Q  & 0.42 &  [n]  &    0.44 & $<$0.9 & J  & VT96 &      & FG85 & HB95 \\
1841+672   &           & 0.470 & G  & 0.16 &  0.15 &    0.05 &    6.4 & C  & SB96 &      &      & Sn97 \\[0.15cm]
1842+681   &           & 0.475 & Q  & 0.93 &  0.61 &    0.34 &    1.9 & C  & VV96 &      &      & XR95 \\
1843+356   &           & 0.764 & G  & 0.79 &  1.03 & $-$0.21 &   10.3 & C  & VT96 &      &      & XR95 \\
1845+797   & 3C 390.3  & 0.056 & G  & 4.41 & 11.23 & $-$0.75 &    4.9 & C  & HB91 & KN81 &      & AW96 \\
1849+670   &           & 0.657 & Q  & 0.85 &  0.90 & $-$0.05 &    2.6 & C  & VV96 &      &      & TV94 \\
1850+402   &           & 2.12  & Q  & 0.53 &  0.55 & $-$0.03 &    2.4 & C  & VV96 &      &      & HB95 \\[0.15cm]
1851+488   &           & 1.250 & Q  & 0.31 &  0.29 &    0.05 & $<$1.6 & L  & VT96 &      &      & TV94 \\
1856+737   &           & 0.460 & Q  & 0.58 &  0.56 &    0.03 &    6.6 & C  & VV96 &      &      & TV94 \\
1901+319   & 3C 395    & 0.635 & Q  & 1.86 &  2.95 & $-$0.37 &   15.7 & C  & VV96 &      &      & SH88 \\
1908+484   &           & 0.513 & Q  & 0.50 &  0.58 & $-$0.12 & $<$1.6 & J  & HB97 &      &      & TV94 \\
1910+375   &           & 1.104 & Q  & 0.41 &  0.50 & $-$0.16 & $<$2.3 & J  & HB97 &      &      & HB95 \\[0.15cm]
1915+657   &           & 0.486 & Q  & 0.35 &  0.77 & $-$0.63 &   32.6 & C  & HB97 &      &      & HB95 \\
1921$-$293 &           & 0.352 & Q  &10.60 &  5.70 &    0.49 &    6.7 & C  & VV96 & WO90 & WO90 & SW97 \\
1924+507   &           & 1.098 & Q  & 0.35 &  0.66 & $-$0.51 &    2.1 & C  & VV96 &      &      & HB95 \\
1928+738   &           & 0.303 & Q  & 3.63 &  3.91 & $-$0.06 &    8.4 & C  & VV96 &      &      & GM95 \\
1936+714   &           & 1.864 & Q  & 0.40 &  0.62 & $-$0.35 & $<$1.1 & J  & VV96 &      &      & TV94 \\[0.15cm]
1943+546   &           & 0.263 & G  & 0.94 &  1.65 & $-$0.45 &   40.9 & C  & SK93 &      &      & XR95 \\
1945+604   &           & 2.700 & Q  & 0.08 &  0.06 &    0.23 & $<$1.0 & J  & Sn97 &      & Sn97 & Sn97 \\
1946+708   &           & 0.101 & G  & 0.68 &  0.92 & $-$0.24 &   33.2 & C  & SK93 &      &      & TV94 \\
1950+573   &           & 0.652 & Q  & 0.51 &  0.57 & $-$0.09 &   13.5 & C  & HB97 &      &      & TV94 \\
1954$-$388 &           & 0.630 & Q  & 2.00 &  1.59 &    0.18 & $<$2.1 & L  & VV96 & WO90 & WO90 & SW98 \\[0.15cm]
1957+405   & Cyg A     & 0.056 & G  & 0.74$^q$ & 0.78$^q$ & $-$0.04 &   15.4 & C  & VV96 & CB94 & CB94 & CB94 \\
1958+619   &           & 1.824 & Q  & 0.14 &  0.11 &    0.19 & $<$1.1 & J  & Sn97 &      & Sn97 & Sn97 \\
2005+642   &           & 1.574 & Q  & 0.72 &  0.17 &    1.16 & $<$1.1 & J  & HB97 &      &      & TV94 \\
2007+659   &           & 1.325 & Q  & 0.75 &  1.03 & $-$0.26 &    2.3 & C  & VV96 &      &      & TV94 \\
2007+777   &           & 0.342 & B  & 1.28 &  0.94 &    0.25 &    3.6 & C  & VV96 & KP81 &      & GM94 \\[0.15cm]
2015+657   &           & 2.845 & Q  & 0.53 &  0.97 & $-$0.49 & $<$1.2 & J  & VV96 &      &      & TV94 \\
2017+745   &           & 2.187 & Q  & 0.54 &  0.47 &    0.11 &    4.9 & C  & HB97 &      &      & TV94 \\
2021+614   &           & 0.227 & G  & 2.62 &  2.13 &    0.17 &    9.1 & C  & VV96 &      &      & PR88 \\
2043+749   & 4C 74.26  & 0.104 & Q  & 0.37 &  1.60 & $-$1.18 &    2.6 & C  & VV96 &      &      & PB92 \\
2048+312   &           & 3.198 & Q  & 0.59 &  0.75 & $-$0.19 &    2.8 & C  & VV96 &      & VL96 & GS94 \\
\noalign{\smallskip}                       
\hline                                     
\end{tabular}                              
                                           
   \end{table*}

\setcounter{table}{0}
%
   \begin{table*}                          
      \caption{{\it continued}}                           
      \label{TSou6}                        
\begin{tabular}{clrcrrrrcllll}                    \hline
      \noalign{\smallskip}                 
                        1      &    
\,\,\,\,\,\,\,\,\,\,    2      &    
3\,\,\,\,                      &    
4                              &    
5\,\,\,\,                      &    
6\,\,\,\,                      &    
7\,\,\,\,                      &    
8\,\,\,\,\,                    &    
9                              &    
\,\,                   10      &    
\,\,                   11      &    
\,\,                   12      &    
\,\,                   13      \\   
      \noalign{\smallskip}                 
      \hline                               
      \noalign{\smallskip}      
2116+818   &           & 0.084 & G  & 0.38 &  0.54 & $-$0.28 &    4.7 & C  & MB96 & KP81 &      & TV96 \\
2121+053   &           & 1.941 & Q  & 2.78 &  1.14 &    0.72 & $<$1.0 & L  & VV96 &      &      & WC92 \\
2134+004   &           & 1.932 & Q  & 9.96 &  3.13 &    0.91 &    4.0 & C  & VV96 & WO90 & WO90 & SW97 \\
2136+141   &           & 2.427 & Q  & 1.07 &  1.15 & $-$0.06 &    2.9 & C  & VV96 &      &      & WC92 \\
2136+824   &           & 2.357 & Q  & 0.51 &  1.01 & $-$0.55 &   13.7 & C  & HB97 & KP81 &      & TV94 \\[0.15cm]
2145+067   &           & 0.999 & Q  & 4.14 &  2.88 &    0.29 &    7.2 & C  & VV96 &      &      & WC92 \\
2155$-$152 &           & 0.672 & Q  & 1.58 &  1.26 &    0.18 &    5.4 & C  & VV96 & WO90 & WO90 & SW90 \\
2200+420   &  BL Lac   & 0.069 & B  & 2.94 &  4.69 & $-$0.38 &    7.0 & C  & VV96 &      &      & PR88 \\
2201+315   &           & 0.298 & Q  & 2.81 &  1.98 &    0.28 &    5.4 & C  & VV96 &      &      & DW87 \\
2207+374   &           & 1.493 & Q  & 0.86 &  1.71 & $-$0.55 &   55.6 & C  & VV96 &      &      & XR95 \\[0.15cm]
2223$-$052 & 3C 446    & 1.404 & Q  & 4.51 &  6.37 & $-$0.26 &    4.9 & C  & VV96 & KN81 &      & WC90 \\
2230+114   & CTA 102   & 1.037 & Q  & 3.97 &  6.63 & $-$0.41 &   15.7 & C  & VV96 &      &      & WC89 \\
2235+731   &           & 1.345 & Q  & 0.39 &  0.30 &    0.21 &    2.6 & C  & VV96 &      &      & TV94 \\
2243$-$123 &           & 0.630 & Q  & 2.38 &  [o]  & $-$0.22 &    1.8 & C  & VV96 & WO90 & WO90 & HV98 \\
2246+370   &           & 1.541 & Q  & 0.43 &  0.91 & $-$0.60 & $<$0.9 & J  & VT96 &      &      & TV94 \\[0.15cm]
2251+134   &           & 0.677 & Q  & 0.88 &  1.44 & $-$0.40 &    1.4 & C  & VV96 &      &      & HS92 \\
2251+158   & 3C 454.3  & 0.859 & Q  &14.47 & 13.90 &    0.03 &    7.8 & C  & VV96 &      &      & CG96 \\
2253+417   &           & 1.476 & Q  & 1.08 &  1.41 & $-$0.21 &    4.1 & C  & VV96 &      &      & XR95 \\
2255+417   & 4C 41.45  & 2.150 & Q  & 1.10 &  1.87 & $-$0.43 &   43.5 & C  & VT96 &      &      & XR95 \\
2259+371   &           & 2.179 & Q  & 0.44 &  0.60 & $-$0.25 &    6.2 & C  & HB97 &      &      & TV94 \\[0.15cm]
2309+454   &           & 1.447 & Q  & 0.50 &  0.31 &    0.38 &    2.0 & C  & VV96 &      &      & TV94 \\
2310+385   &           & 2.181 & Q  & 0.53 &  0.69 & $-$0.21 &    4.4 & C  & HB97 &      &      & HB95 \\
2330+387   &           & 0.319 & Q  & 0.43 &  0.84 & $-$0.54 &   46.2 & C  & VV96 &      &      & HB95 \\
2335+267   & 3C 465    & 0.029 & G  & 1.56 &  7.46 & $-$1.26 &    8.1 & C  & DV91 &      &      & VC95 \\
2345$-$167 &           & 0.576 & Q  & 3.47 &  1.20 &    0.85 &    2.5 & C  & VV96 & WO90 & WO90 & HV98 \\[0.15cm]
2351$-$154 &           & 2.675 & Q  & 0.93 &   [p] & $-$0.24 & $<$1.2 & L  & VV96 & WO90 & WO90 & HV98 \\
2352+495   &           & 0.237 & G  & 1.60 &  2.34 & $-$0.31 &   32.8 & C  & CP92 &      &      & CP92 \\
2353+816   &           & 1.344 & B  & 0.48 &  0.40 &    0.15 &    1.7 & C  & VT96 & KP81 &      & TV94 \\ 
2355$-$534 &           & 1.006 & Q  & 1.66 &  1.22 &    0.24 &    4.9 & C  & VV96 & WO90 & WO90 & SW98 \\
2356+390   &           & 1.201 & Q  & 0.36 &  0.43 & $-$0.14 &    8.8 & C  & HB97 &      &      & HB95 \\
\noalign{\smallskip}                                                                                      
\hline                                                                                                    
\end{tabular}                                                                                             
   \begin{list}{}{}
\item $^a$ Optical counterpart: Q -- quasar, B -- BL Lac object, G --
radio galaxy (including Seyfert galaxies of all types).
\item$^b$ Characteristic angular size (case C) or its upper limit (cases
          J, L, and S). 
\item $^c$ Radio structure code:
   \begin{itemize}
   \item C -- size between the core and the most distant 2\%-component;
   \item J -- upper limit of $\theta$ along jet direction;
   \item L -- upper limit of $\theta$ along major axis of the beam;
   \item S -- upper limit of $\theta$ along minor axis of the beam.
   \end{itemize}
\item $^d$ 6 cm flux density from Gregory et al. (1996) unless otherwise stated. 
\item $^e$ Flux density from White and Becker (1992) unless otherwise
stated, at 20 cm or at an alternative wavelength as stated in footnotes
[f -- y].
\item $^f$ Spectral index calculated using $S_{11} = 0.61$ Jy.
\item $^g$ Spectral index calculated using $S_{11} = 3.56$ Jy.
\item $^h$ Spectral index calculated using $S_{75} = 5.05$ Jy.
\item $^i$ Spectral index calculated using $S_{11} = 1.10$ Jy.
\item $^j$ Spectral index calculated using $S_{11} = 1.33$ Jy.
\item $^k$ Spectral index calculated using $S_{11} = 0.40$ Jy.
\item $^l$ Spectral index calculated using $S_{11} = 0.56$ Jy.
\item $^m$ Spectral index calculated using $S_{11} = 0.45$ Jy.
\item $^n$ Spectral index calculated using $S_{75} = 0.14$ Jy.
\item $^o$ Spectral index calculated using $S_{11} = 2.74$ Jy.
\item $^p$ Spectral index calculated using $S_{11} = 1.08$ Jy.
\item $^q$ $S_6$ and $S_{20}$ values for the compact core component.
\item $^x$ Spectral index calculated using $S_{11} = 1.76$ Jy.
\item $^y$ Spectral index calculated using $S_{11} = 1.34$ Jy
   \end{list}

\end{table*}
   

\setcounter{table}{0}
%
   \begin{table*}                          
      \caption{{\it continued}: List of references.}                           
      \label{TSou7}                        
\begin{tabbing}
KP81 -- 11 cm data from K\"{u}hr 1981***  \= 
KP81 -- 11 cm data from K\"{u}hr 1981***  \=
KP81 -- 11 cm data from K\"{u}hr 1981***  \kill
AA92 -- Aller et al. 1992              \>
AW96 -- Alef et al. 1996               \> 
BS95 -- Brinkman et al. 1995           \\ 
CB94 -- Carilli et al. 1994            \> 
CB96 -- Chu et al. 1996                \>
CG96 -- Cawthorne and Gabuzda 1996     \\
CP92 -- Conway et al. 1992             \>
CW93 -- Cawthorne et al. 1993          \>
DS93 -- Downes et al. 1993             \\ 
DV91 -- De Vacoulers G.  et al. 1991   \>
DW87 -- De Waard G. 1987               \>
FC93 -- Feretti et al. 1993            \\
FG85 -- Ficarra et al. 1985            \>  
FG97 -- Frey et al. 1997               \> 
GC92 -- Gabuzda et al. 1992            \\
GF94 -- Giovannni et al. 1994          \>
GK92 -- Gurvits et al. 1992            \>
GM94 -- Gabuzda et al. 1994            \\
GM95 -- Guirado et al. 1995            \>
GS94 -- Gurvits et al. 1994            \>
GV94 -- Gregory et al. 1994            \\
GW89 -- Gabuzda et al. 1989            \>
HB91 -- Hewitt and Burbidge 1991       \>
HB93 -- Hewitt and Burbidge 1993       \\
HB95 -- Henstock et al. 1995           \>
HB97 -- Henstock et al. 1997           \>
HM97 -- Hook and McMahon 1997          \\
HO92 -- Hooimeyer et al. 1992a         \> 
HR92 -- Herbig and Readhead 1992       \>
HS92 -- Hooimeyer et al. 1992b         \\
HV98 -- Hong et al. 1998               \>
JU86 -- Jones et al. 1986              \>
Ke85 -- Keel 1985                      \\
KN81 -- K\"{u}hr et al. 1981a          \>
KP81 -- K\"{u}hr et al. 1981b          \>
KW90 -- Kus et al. 1990                \\
Lo96 -- Lobanov 1996                   \>
Ma88 -- Marscher 1988                  \>
MB96 -- Marcha et al. 1996             \\
MH88 -- Michel and Huchra 1988         \>
MH96 -- Marzke et al. 1996             \>
MH98 -- McMahon and Hook 1998          \\
MM90 -- McHardy et al. 1990            \>
OL95 -- Owen et al. 1995               \>
OW95 -- Oren and Wolfe 1995            \\
PA90 -- Preuss et al. 1990             \>
PB92 -- Pearson et al. 1992            \>
PC96 -- Perlman et al. 1996            \\
PF98 -- Paragi et al. 1998             \> 
PR88 -- Pearson and Readhead 1988      \>
PW93 -- Polatidis et al. 1993          \\
RB95 -- Romney et al. 1995             \>
Ru88 -- Rusk 1988                      \>
SB96 -- Snellen et al. 1996            \\
SH88 -- Simon et al. 1988              \>
SH92 -- Strauss et al. 1992            \>   
SK93 -- Stickel and K\"{u}hr 1993      \\
SK94 -- Stickel and K\"{u}hr 1994      \> 
Sn97 -- Snellen 1997                   \>
SQ96 -- Standke et al. 1996            \\
SW97 -- Shen et al. 1997               \> 
SW98 -- Shen et al. 1998               \>
TE96 -- Tingay et al. 1996             \\
TM98 -- Tingay et al. 1998             \>
TV94 -- Taylor et al. 1994             \> 
TV96 -- Taylor et al. 1996             \\
UC89 -- Unwin et al. 1989              \>
UT97 -- Udomprasert et al. 1997        \>
VC95 -- Venturi et al. 1995            \\
VG93 -- Venturi et al. 1993a           \>
VL96 -- VLA Calibrator Manual 1996     \>
VP93 -- Venturi et al. 1993b           \\
VT96 -- Vermeulen et al. 1996          \>
VV96 -- Veron-Cetty and Veron 1996     \>
WB87 -- Walker et al. 1987             \\
WB97 -- White et al. 1997              \>
WC89 -- Wehrle and Cohen 1989          \>
WC90 -- Wehrle et al. 1990             \\
WC92 -- Wehrle et al. 1992             \>
WO90 -- Wright and Otrupcek 1990       \>
WS88 -- Witzel et al. 1988             \\
XL94 -- Xu et al. 1994                 \>
XR95 -- Xu et al. 1995                 \>
ZB88 -- Zensus et al. 1988             \\
ZB94 -- Zhang et al. 1994              \>
                                       \>
                                       \\
\end{tabbing}

\end{table*}

   \begin{table*}
\caption{Two-parameter ($lh$ and $q_{0}$) regression model results with
1$\sigma$ errors for different fixed values of $\beta$ and $n$ for the
sample of 145 sources ($Lh^{2} \le 10^{26}$ W/Hz, $-0.38 \le \alpha \le
0.18$).
        }
         \label{Tregr}
         \begin{tabular}{rrrrrrrrr}
            \hline
            \noalign{\smallskip}
$n$                 &
                    & 
$\beta$=$-$0.20     & 
$\beta$=$-$0.10     & 
$\beta$=$-$0.05     & 
$\beta$=0.0         & 
$\beta$=0.05        &   
$\beta$=0.10        & 
$\beta$=0.20        \\
            \noalign{\smallskip}
            \hline
            \noalign{\smallskip}
$-$0.3              & 
$lh$ (pc)           & 
13.98$\pm$4.71      & 
16.48$\pm$4.82      & 
17.56$\pm$5.22      & 
18.48$\pm$6.79      & 
19.20$\pm$9.09      & 
19.90$\pm$1.81      & 
20.84$\pm$2.54      \\
                    &
$q_{0}$             & 
{\bf 1.78$\pm$0.83} & 
{\bf 1.04$\pm$0.51} & 
{\bf 0.81$\pm$0.51} & 
{\bf 0.64$\pm$0.73} & 
{\bf 0.51$\pm$1.59} & 
{\bf 0.41$\pm$0.27} & 
{\bf 0.26$\pm$0.06} \\[0.15cm]

$-$0.2              & 
$lh$ (pc)           & 
14.64$\pm$4.03      & 
16.76$\pm$4.14      & 
17.68$\pm$4.50      & 
18.42$\pm$6.07      & 
19.00$\pm$4.02      & 
19.60$\pm$1.93      & 
20.28$\pm$2.44      \\
                    & 
$q_{0}$             & 
{\bf 1.22$\pm$0.43} & 
{\bf 0.73$\pm$0.30} & 
{\bf 0.57$\pm$0.32} & 
{\bf 0.45$\pm$0.53} & 
{\bf 0.36$\pm$0.71} & 
{\bf 0.28$\pm$0.15} & 
{\bf 0.17$\pm$0.03} \\[0.15cm]

$-$0.1              & 
$lh$ (pc)           & 
15.02$\pm$3.48      & 
16.82$\pm$3.58      & 
17.58$\pm$3.87      & 
18.16$\pm$5.42      & 
18.70$\pm$2.06      & 
19.00$\pm$1.98      & 
19.60$\pm$2.30      \\
                    & 
$q_{0}$             & 
{\bf 0.86$\pm$0.23} & 
{\bf 0.52$\pm$0.18} & 
{\bf 0.40$\pm$0.20} & 
{\bf 0.31$\pm$0.40} & 
{\bf 0.24$\pm$0.33} & 
{\bf 0.19$\pm$0.09} & 
{\bf 0.10$\pm$0.02} \\[0.15cm]

0.0                 & 
$lh$ (pc)           & 
15.15$\pm$3.04      & 
16.68$\pm$3.12      & 
17.22$\pm$3.36      & 
17.72$\pm$4.83      & 
18.10$\pm$1.62      & 
18.33$\pm$1.95      & 
18.66$\pm$2.14      \\
                    & 
$q_{0}$             & 
{\bf 0.60$\pm$0.13} & 
{\bf 0.36$\pm$0.11} & 
{\bf 0.28$\pm$0.13} & 
{\bf 0.21$\pm$0.30} & 
{\bf 0.16$\pm$0.18} & 
{\bf 0.12$\pm$0.05} & 
{\bf 0.05$\pm$0.01} \\[0.15cm]

0.1                 & 
$lh$ (pc)           & 
15.14$\pm$2.69      & 
16.36$\pm$2.74      & 
16.80$\pm$2.92      & 
17.14$\pm$4.30      & 
17.30$\pm$1.56      & 
17.56$\pm$1.89      & 
17.68$\pm$1.99      \\
                    & 
$q_{0}$             & 
{\bf 0.42$\pm$0.08} & 
{\bf 0.25$\pm$0.07} & 
{\bf 0.19$\pm$0.09} & 
{\bf 0.14$\pm$0.23} & 
{\bf 0.10$\pm$0.11} & 
{\bf 0.07$\pm$0.03} & 
{\bf 0.01$\pm$0.01} \\[0.15cm]

0.2                 & 
$lh$ (pc)           & 
14.92$\pm$2.40      & 
15.88$\pm$2.43      & 
16.20$\pm$2.56      & 
16.46$\pm$3.81      & 
16.60$\pm$1.55      & 
16.74$\pm$1.79      & 
16.24$\pm$2.05      \\
                    & 
$q_{0}$             & 
{\bf 0.29$\pm$0.05} & 
{\bf 0.16$\pm$0.05} & 
{\bf 0.12$\pm$0.06} & 
{\bf 0.08$\pm$0.18} & 
{\bf 0.05$\pm$0.07} & 
{\bf 0.02$\pm$0.02} & 
{\bf 2e-6$\pm$1e-3} \\[0.15cm]

0.3                 & 
$lh$ (pc)           & 
14.52$\pm$2.16      & 
15.30$\pm$2.16      & 
15.54$\pm$2.26      & 
15.69$\pm$3.38      & 
15.80$\pm$1.53      &
15.40$\pm$1.65      &
14.63$\pm$1.29      \\
                    &
$q_{0}$             & 
{\bf 0.20$\pm$0.03} & 
{\bf 0.10$\pm$0.03} & 
{\bf 0.06$\pm$0.04} & 
{\bf 0.03$\pm$0.15} &
{\bf 7e-3$\pm$0.04} &
{\bf 5e-3$\pm$0.02} &
{\bf 1e-7$\pm$1e-6} \\
            \noalign{\smallskip}
            \hline
         \end{tabular}
   \end{table*}

\cleardoublepage

\end{document}